\documentclass[prb,aps,amsmath,amssymb,showpacs,twocolumn,superscriptaddress]{revtex4}
\usepackage{graphicx}
\usepackage{color}

\newcommand{\beq}{\begin{equation}}
\newcommand{\eeq}{\end{equation}}
\newcommand{\bma}{\begin{math}}
\newcommand{\ema}{\end{math}}
\newcommand{\beqa}{\begin{eqnarray}}
\newcommand{\eeqa}{\end{eqnarray}}
\def\opone{\le\textbf{}\textbf{}avevmode\hbox{\small1\kern-3.8pt\normalsize1}}

\def\da{\downarrow}
\def\ua{\uparrow}

\newcommand {\be}[1]{
{\marginpar{{\scriptsize\ \\ \ #1}}}
\begin{eqnarray} \mbox{$\label{#1}$} }

\newcommand{\ee}{\end{eqnarray}}

\newcommand\ket [1] {|#1 \rangle }
\newcommand\bra [1] {\langle #1 |}

\newcommand{\ul}\underline

\newcommand{\cM}{\mathcal{M}}
\newcommand{\cO}{\mathcal{O}}
\newcommand{\cS}{\mathcal{S}}
\newcommand{\ba}{{\bf a}}
\newcommand{\bb}{{\bf b}}

\DeclareMathOperator*{\pf}{Pf}

\DeclareMathOperator{\Det}{Det}
\DeclareMathOperator{\sld}{sl}

\allowdisplaybreaks[4]

\newlength{\cwidth}

\newlength{\xw}
\begin{document}

\title{The structure of spinful quantum Hall states: a squeezing perspective}

\author{E. Ardonne}
\affiliation{Nordita, Roslagstullsbacken 23, SE-106 91 Stockholm, Sweden}
\author{N. Regnault} 
\affiliation{Laboratoire Pierre Aigrain, ENS and CNRS, 24 rue Lhomond, 75005 Paris, France}

\date{November 1 , 2011}


\begin{abstract}
We provide a set of rules to define several spinful quantum Hall model states. The method extends the one that is known for spin-polarized states. It is achieved by specifying an undressed root partition, a squeezing procedure, and rules to dress the configurations with spin. It applies to both the excitationless  and the quasihole states. In particular, we show that the naive generalization where one preserves the spin information during the squeezing sequence, may fail. We give numerous examples such as the Halperin states, the non-abelian spin singlet states, or the spin-charge separated states. The squeezing procedure for the series $(k=2,r)$ of spinless quantum Hall
states, which vanish as $r$ powers when $k+1$ particles coincide, is generalized to the
spinful case. As an application of our method, we show that the counting observed in the particle entanglement spectrum of several spinful states matches the one obtained through the root partitions and our rules. This counting also matches the counting of quasihole states of the
corresponding model Hamiltonians, when the latter are available.
\end{abstract}

\maketitle

\section{Introduction}

The theoretical study of the fractional quantum Hall (FQH) effect has relied on model wave functions since its discovery \cite{laughlin83}. They provide an easy way to understand the physical properties of an inherently hard quantum $n$-body problem. In addition, having knowledge of the wave functions
representing different topological phases provides insight to the question as to which topological
phases can exist. Trying to fully classify all topological phases is a tremendous task, but progress
has been made in the context of topological insulators and superconductors
(see, for instance, \onlinecite{srf09,k09}).

In the context of the quantum Hall wave functions, progress also has been made in
several ways. A popular and successful approach has been to study model Hamiltonians,
in combination with conformal-field-theory techniques. In this approach, one studies the
zero-energy ground states of model electron-electron interactions, which gives rise
to certain vanishing properties of the model quantum Hall wave functions. The simplest
example of this is the Laughlin wave function (say, at filling $\nu=1/3$), which is the
unique, densest zero-energy ground state of the model interaction given by the
Haldane pseudopotential \cite{h83}. Excitations of quantum Hall states can be
created by changing the flux. Upon increasing the flux, one creates quasihole
states, which still are zero-energy ground states of the model Hamiltonian. These
quasiholes can have fractional charge \cite{laughlin83},
fractional statistics \cite{asw84}, and even non-abelian
statistics, which was pioneered in \onlinecite{mr91}. For recent developments
with regard to the non-abelian Berry phase, we refer to
\onlinecite{r09,bgn11}.  

The model Hamiltonians, for which the quasihole states are the exact ground states,
only constrain the behaviour of the underlying electrons. Thus, it should be possible
to infer the properties of the anyons from the properties of the electrons in the quantum
liquid alone. This implies that, for model quantum Hall states, there should be a
`duality' between the electrons and excitations. Such a `duality' has been observed a
long time ago already. In the context of the low-energy Chern-Simons description of
the abelian Laughlin states, we refer to \onlinecite{w95} (see also \onlinecite{es98} for a detailed
account on the edge state version of this duality). Subsequently, this notion has been
extended to non-abelian quantum Hall states (see for instance \onlinecite{abs01}).
More recently, a seemingly related duality was observed between the conformal field theory
correlators describing the electron and quasihole states \cite{bbs10up}.

The notion of this duality is important, because it implies that it should be possible to
deduce the properties of the excitations from the ground-state wave functions, and
therefore restricts the number of wave functions that can describe topological phases.
In addition, it puts constraints on the underlying (conformal) field theory description of
topological phases in the quantum Hall effect setting. Apart from this duality, there is
another, more practical, constraint that one can impose, namely, requiring that the
wave functions one considers are eigenstates of a local model Hamiltonian. This constraint
allows one to effectively study the topological properties of the state, and provides a way of
uniquely defining (or specifying) the state, by a small set of rules. We note that, for instance,
the successful Jain states \cite{j89} do not satisfy this constraint, so this is not a physical,
but rather a practical requirement, to obtain a more tractable, but still very rich and
interesting problem. We also stress that even if the state satisfies the duality and is a ground
state of a local Hamiltonian, this does not imply that the wave function represents a
genuine topological phase of matter \cite{r09}.

Recent developments in generating candidate quantum Hall wave functions gave rise to a framework based on root partitions, squeezing, and highest weight conditions that provides an elegant manner to address several candidates wave functions. This includes the ground state, its quasihole \cite{bh08b,bh08c} and some aspects of quasielectron excitations
as well as excitons \cite{bh09} (see also \onlinecite{jj03,hhr09,hhs09,rxxup} for
more details on quasielectrons and excitions).

For the time being, the effort has mostly concentrated on the spin polarized systems. However, spinful FQH states are relevant in many realistic cases. The additional spin degree of freedom can be the true spin of the electrons, a layer index in bilayer systems, pseudospin to handle valley degeneracy, or spin-$1/2$ rotating ultracold fermions. With the success of root partitions for spinless (or spin-polarized) systems, it is worth analyzing how this concept can be translated to the spinful case.   

The main goal of this paper is to give a set of rules, which can be used
to define model quantum Hall states, with spin (or another internal degree
of freedom), by specifying a so-called `root partition', and a squeezing
procedure, which is used to define a Hilbert space. The model states
are then obtained by imposing highest weight conditions on this
Hilbert space, for both the orbital and spin part. The model states we are considering in this paper, can
be uniquely defined in this way, in the case when no excitations are
present. For such a procedure to be meaningful, this procedure should
also work when (quasihole) excitations are present, i.e. in the case
when the number of flux quanta is increased, in comparison to the
state without excitations.

It is not a priori clear how to generalize the
squeezing procedure from polarized states to model states with
spin (or other `internal' degrees of freedom). There are, in principle,
several routes that one might take, but we found that only one of them
correctly generates all the ground states of the model Hamiltonians,
including the quasiholes states. Prior work\cite{reg08,mjv09,hrb11,seidel2011}
has mostly focused on the 
Halperin\cite{halperin83} or Haldane-Rezayi\cite{hr88} states. We show that this concept can be extended to other known states 
such as the non-abelian spin singlet or spin-charge separation states, but can also provide a way to obtain new interesting states.

The outline of the paper is as follows. In section \ref{sec:polarized}, we review the squeezing
procedure for the case of spin polarized quantum Hall states. In section \ref{sec:spinful}, we explain how the root partitions and squeezing technique can be extended to the spinful wave functions. We give several examples in section \ref{sec:singlets}. Interesting series of root partitions is described in section \ref{sec:rootrelations}. It generalizes the spinless series $(k=2,r)$, which include the Moore-Read, Gaffnian and Haffnian states. As an application of our results, we then show in section \ref{sec:applications} that the counting we have obtained for the quasihole excitations matches the counting deduced from the particle entanglement spectrum.

In appendix \ref{app:raising-lowering}, we collect the requirements for a state to be a spin singlet
state, and give the spin-raising and -lowering operators explicitly.
Appendix \ref{app:numerics} briefly
describes how the various highest weight conditions can be implemented on the (reduced)
Hilbert spaces. Finally, in appendix \ref{app:counting}, we collect the formulas from the
literature giving the number of quasihole states for a set of model Hamiltonians that we consider
in this paper.

\section{Overview of squeezing for polarized quantum Hall states}
\label{sec:polarized}

Because of the importance of the quasihole states in the spinful case,
it seems prudent to review the spin-polarized case, and pay special
attention to the model state in the presence of quasihole excitations.
Moreover, many of the spin-polarized states can be viewed as
particular spinful states with quasihole excitations.

We focus our attention to those states that have a ground
state than can be uniquely defined by a squeezing procedure,
including the Laughlin \cite{laughlin83}, Moore-Read \cite{mr91} and
Read-Rezayi \cite{readrezayi2} states, as well as for instance the
`Gaffnian' \cite{ymg88,src07} and `Haffnian' \cite{thesis:green01} wave functions.

Quantum Hall states in the lowest Landau level are, apart from
a geometry-dependent `confining factor', given by (anti-) symmetric
polynomials in terms of the coordinates of the (fermionic) bosonic
constituent particles. For simplicity, we will be mainly considering
bosonic states in this paper; fermionic versions can trivially be
obtained by multiplying with an additional global Jastrow factor.
An exception to this rule will be fermionic states which do not
contain a Jastrow factor of all particles, and these states can thus
not be made bosonic by removing an overall Jastrow factor.

Let us now start by reviewing the squeezing procedure for polarized
bosonic quantum Hall states.
Being symmetric polynomials, these states can be expanded in so-called
`symmetrized monomials'. Symmetrized monomials
are labeled by `partitions', or, equivalently, and perhaps more
appropriate in the context of quantum Hall states, `orbital occupation
numbers'. 

To be explicit, let us consider the orbital occupation
$(n_0,n_1,\ldots, n_{N_\phi})$, such that the $l^{\rm th}$ orbital
is occupied with $n_l$ particles. The total number of flux
quanta is denoted by $N_\phi$ (we only consider the spherical
geometry in this paper), while the total number of
particles is given by $N = \sum_{l=0}^{N_\phi} n_l$. The total
degree of the symmetrized monomial corresponding
to these orbital occupation numbers is $d=\sum_{l=0}^{N_\phi} l \,  n_l$.
The partition $\mu$ partitions the total degree $d$ and has
$n_l$ rows of length $l$. As an example, we take the orbital
occupation $(2,0,2)$, which corresponds to a symmetrized
monomial of total degree four. The corresponding partition is
$\mu = (2,2,0,0)$, where we included the zeros, which
indicate that the zero'th orbital is doubly occupied. In addition,
including the zeros ensures that the length of the vector
describing the partition equals the number of particles.
The `elements' of the partition $\mu$ will denoted
by $\mu_i$. Now, the symmetrized monomials 
$m_\mu$ corresponding to $\mu$ is given by
\begin{equation}
m_\mu = \cS \bigl[ 
z_1^{\mu_1} z_2^{\mu_2} \cdots z_N^{\mu_N}
\bigr] \ ,
\end{equation} 
where $z_i$ is the complex coordinate of the $i^{\rm th}$ particle and 
$\cS$ denotes the symmetrization, which 
is normalized such that each term in the symmetrization has coefficient one.
In particular, in the case of the partitions $\mu = (2,2,0,0)$,
$\mu = (2,1,1,0)$ and $\mu = (1,1,1,1)$, corresponding
to the orbital occupations $(2,0,2)$, $(1,2,1)$ and $(0,4,0)$
respectively, one obtains
\begin{align*}
m_{(2,2,0,0)} &= z_1^2 z_2^2 + z_1^2 z_3^2 + z_1^2 z_4^2
+ z_2^2 z_3^2 + z_2^2 z_4^2 + z_3^2 z_4^2 \\
m_{(2,1,1,0)} &=
z_1^2 z_2 z_3 + z_1 z_2^2 z_3 + z_1 z_2 z_3^2 + z_1^2 z_2 z_4 \\
& + z_1 z_2^2 z_4 + z_1 z_2 z_4^2 + z_1^2 z_3 z_4 + z_1 z_3^2 z_4\\
& + z_1 z_3 z_4^2 + z_2^2 z_3 z_4 + z_2 z_3^2 z_4 + z_2 z_3 z_4^2 \\
m_{(1,1,1,1)} &=  z_1 z_2 z_3 z_4
\end{align*}

As stated above, any symmetric polynomial in a certain number of variables can
be expressed in terms of symmetrized monomials,
\begin{equation}
\Psi_{\rm sym} (\{ z_i \}) = \sum_{\mu} c_\mu m_\mu (\{z_i\}) \ .
\end{equation}
For comparison, anti-symmetric wave functions describing fermions
can be expanded in anti-symmetric monomials (i.e., Slater determinants), which
are written as $\sld_\mu = \Det \bigl( z_i^{\mu_j} \bigr)$.

Inspired by quantum Hall states on a spherical geometry \cite{h83}, we
assign an orbital angular momentum $l_z$ to each orbital. We choose
the convention that the $l_z$ quantum numbers of the orbitals are
given by $(-N_{\phi}/2,-N_\phi/2+1,\ldots,N_\phi /2 - 1,N_\phi /2)$, i.e., the
orbital corresponding to $z^0$ has the lowest angular momentum
$-N_\phi/2$. With this convention, we have that the angular momentum
operators are given by
\begin{align}
L_- & = \sum_{i=1}^{N} \partial_{z_i} & 
L_z & = N N_\phi/2 - \sum_{i=1}^{N} z_i \partial_{z_i} \nonumber \\ 
L_+ & = \sum_{i=1}^{N} N_\phi z_i -z_i^2 \partial_{z_i}.
\end{align}

With these preliminaries in place, we can now explain how various
model states can be completely specified by a few simple rules. First, for all
states, there is a unique `highest' symmetrized monomial. The concept
of `highest' can be defined in a few different, but equivalent ways. In terms
of the orbital occupation numbers, all the orbital occupation numbers of the
symmetric monomials present in the expansion of the states can be obtained
from the highest one, by `squeezing' particles inwards (such that
all the symmetrized monomials have the same angular momentum).
In terms of the orbital occupation numbers, the squeezing process takes the
following form. Taking two particles (assumed to be bosons for now)
in orbitals $i$ and $j$, with $i<j-1$, we move
these particles to orbitals $i+1$ and $j-1$ respectively. Explicitly, if one starts
with the orbital occupation
$(n_0,n_1,\ldots,n_{i},n_{i+1},\ldots,n_{j-1},n_{j},\ldots,n_{N_\phi})$, one ends up
with
$(n_0,n_1,\ldots,n_{i}-1,n_{i+1}+1,\ldots,n_{j-1}+1,n_{j}-1,\ldots,n_{N_\phi})$,
after squeezing particles in orbitals $i$ and $j$. In the case of spinless
fermions, one need that $i<j-2$, as well as $n_{i+1} = n_{j-1} = 0$, because
of the Pauli principle.

It was realized by Haldane and Bernevig \cite{bh08a,bh08b}, that many model
quantum Hall states can be written as a single Jack polynomial (with negative
parameter). Such Jack polynomials had been studied in the literature \cite{fjm02},
and indeed have a highest root configuration.

Alternatively, this highest orbital occupation (at least for states in the absence of
quasihole excitations) also corresponds to that part of the wave function
that survives if one puts the wave function on the cylinder, and takes the thin-cylinder
(or Tao-Thoules) limit \cite{bk05,bkw06,sl06}.
In this limit, only those states that maximize $\sum_{i=0}^{N_\phi} n_i^2$, survive.

Finally, in mathematical terms, one says that the highest partition `dominates'
all the partitions corresponding to symmetrized monomials present in the wave function.
A partition $\mu$ dominates a partition $\lambda$ if $\lambda$ can be obtained
from $\mu$ by successive squeezing operations on $\mu$.
If there exists a highest dominating partition (which is not completely un-squeezed),
one can reduce the sum over partitions $\mu$ in the expansion in terms of
symmetrized monomials over those partitions $\mu$ which are dominated by the
root partition $\lambda$, which is denoted as $\mu \leq \lambda$,
\begin{equation}
\Psi (\{ z_i \}) = \sum_{\mu\leq \lambda} c_\mu m_\mu (\{z_i\}) \ .
\end{equation}

The existence of a dominating partition, which is `smaller' than the completely un-squeezed
partition, means that one can define a reduced Hilbert
space, by taking this highest partition, and obtaining all the states in the reduced Hilbert
space by using the squeezing operation successively. In general, this reduced Hilbert space is
significantly smaller than the full Hilbert space, which can be exploited in explicit
calculations.

We will now explain how one can completely specify a large set of model quantum
Hall states, first in the case when no quasihole excitations are present, and then in the
presence of quasiholes. The starting point is the model Hamiltonian, which will
be used to obtain the highest, or root partition. For the (bosonic) Read-Rezayi states\cite{readrezayi2},
which we will take as an example throughout this section,
the model Hamiltonian simply gives a positive energy any time $k+1$ particles are
coincident. In the thin-cylinder limit, this interaction translates to an assignment of a
positive energy every time two neighboring angular momentum orbitals have a
total occupation that is bigger than $k$. One can show, for instance, by an explicit
calculation using the thin-cylinder limit, that the root-partition, in the absence of
quasihole excitations, is given by $(k,0,k,\ldots,0,k)$, see \cite{abk08} and \cite{bh08b}.
To obtain the full Read-Rezayi state, one constructs the (reduced) Hilbert space,
which contains all those symmetrized monomials, which can be obtained from the
one with root partition $(k,0,k,\ldots,0,k)$ by symmetrically squeezing inwards. 
Total angular momentum is a good quantum number, and because we are looking
at a state without excitations, we have an $L=0$ state. Note that indeed, the root
partition has $L_z = 0$. To obtain the $L=0$ states in the reduced Hilbert space,
one needs to impose the condition $L_+ \Psi = 0$. Because the Read-Rezayi state
is the unique, highest-density state that vanishes when $k+1$ particles come
together, it is ensured that this procedure will completely determine the coefficients
of the monomials that form a basis for the reduced Hilbert space.  

We will now focus on constructing the Read-Rezayi states in the presence of
quasiholes. We will only be interested in those parts of the wave functions which
depend on the coordinates of the underlying particles. Increasing the flux will lead
to the introduction of quasihole excitations. The ground state of the model interaction
will, in general, be degenerate. For the abelian Laughlin state, there is a so-called
orbital degeneracy of the quasihole excitations. In the case of the non-abelian
quantum Hall states, there is an additional `intrinsic' degeneracy, coming from the
non-abelian nature of the states. This degeneracy is present, even when the
quasiholes are completely localized (in which case the orbital degeneracy is
absent). We should note that the explicit counting of states in the presence of
quasihole excitations has been studied extensively, resulting in an
explicit counting formula, from which all the angular momentum multiplets can
easily be obtained. For details on these counting formulas for
various model states, we refer to \onlinecite{rr96,gr00,arr01,a02}.

To determine which angular momentum states are present for
a given number of quasiholes, and to obtain these states explicitly,
one can also use the above squeezing procedure. First, one needs to
obtain the root partitions for the various angular momenta $L_z$. The highest
angular momentum is of course obtained by `shoving' all particles as far as
possible to the highest angular momentum orbital as possible, i.e. as allowed by
the model Hamiltonian. This will by
definition also be the state which survives in the thin-cylinder limit. The root
partitions of the states at lower angular momentum are obtained by successively
moving a (or the) particle with the lowest angular momentum to lower angular
momenta. Once this particle is in the lowest angular momentum, one takes
the next particle that has lowest angular momentum, and moves it to lower
angular momenta (of course, in such a way that one does not violate the
interaction). One stops with this procedure
when, in the following step, one would obtain a state with negative $L_z$.
By construction, the states described above will survive the thin-cylinder
limit (within the corresponding $L_z$ sector). As an example, we will look at the
$\nu=1$ bosonic Moore-Read state for six particles, with $\Delta N_\phi=2$
added flux quanta, or four quasiholes. In this case, one has the following
root partitions for the various angular momenta:
\begin{align*}
(0,0,2,0,2,0,2) & \quad L_z = 6 \\
(0,1,1,0,2,0,2) & \quad L_z = 5 \\
(1,0,1,0,2,0,2) & \quad L_z = 4 \\
(1,1,0,0,2,0,2) & \quad L_z = 3 \\
(2,0,0,0,2,0,2) & \quad L_z = 2 \\
(2,0,0,1,1,0,2) & \quad L_z = 1 \\
(2,0,1,0,1,0,2) & \quad L_z = 0 \ .
\end{align*}

To determine the number of
multiplets with $L = l$, one takes the root partition corresponding to this
sector, and constructs the associated reduced Hilbert space by squeezing.
On this Hilbert space, one acts with the constraint $L_+ \Psi = 0$. This will
give a set of equations on the coefficients of the basis states. The number of
non-trivial solutions of this set of equations is the number of $L=l$ angular
momentum states. After having obtained these highest weight states, with
$L_z = l$, it is a simple matter to obtain the other states in the same multiplet
by acting with $L_-$.

\section{Squeezing rules in the multi-component case}
\label{sec:spinful}

After having reviewed the squeezing rules in the one-component,
spin-polarized case, we now turn our attention to the main topic of
this paper, the squeezing rules in the multi-component case. The
additional degree of freedom could, for instance, be a layer or valley
degree of freedom, but in this paper, we will focus on spin-1/2 particles.
Of course, the considerations apply to more general, multi component
states as well.

As in the spinless, or polarized case, we will be concerned with model Hamiltonians,
which have a unique, zero-energy ground state, in the absence of
quasihole excitations. 

In the case of bosonic states, the quantum Hall states are symmetric in the
coordinates of the several components separately, which implies that
we can expand them in the following way in terms of symmetrized monomials
\begin{equation}
\label{eq:spin-expansion}
\Psi (\{ z^\ua_i,z^\da_j\} ) = \sum_{\mu ,\mu'} c_{\mu,\mu'} m_{\mu} (\{z^\ua_i\}) m_{\mu'} (\{z^\da_j\}) \ ,
\end{equation}
where the $z^\ua_i$ and $z^\da_j$ are the coordinates of the spin-up and -down
particles, respectively, and the $c_{\mu,\mu'}$ are coefficients. For fermionic
states, one writes the states in terms of Slater determinants instead. The main
point that we address in this section is how one can use squeezing to
reduce the Hilbert space (i.e., to identify a large class of coefficients $c_{\mu,\mu'}$
in the expansion in eq. \eqref{eq:spin-expansion} which are zero).
Equipped with this reduced Hilbert space,
we will again (as in the polarized, one-component case), explain how to explicitly
obtain the various quasihole states (and, hence, also the number of quasihole states
present at a given flux).

\subsection{Some considerations about root configurations}

The main objective will be to find a generalization of the
squeezing rules of the polarized case outlined in the previous section to the
spinful case, for several model states.

In the previous section, we explained that, for the ground state (i.e., in the absence of (quasihole)
excitations) for several model states, there
is a {\em unique} partition, the root partition, from which all the other basis states could
be obtained by successively `squeezing' particles inward in all possible ways. This
root configuration was identical to the unique root configuration of the ground state, which
survived in the thin-cylinder limit.

In the spinful cases, there are, in general, several configurations that survive
the thin-cylinder limit, because this limit is insensitive to the spin (or other internal)
degrees of freedom. Some of these configurations might be forced to have zero
coefficient, due to the explicit form of the model Hamiltonian. Thus, it is not
a priori clear how to generalize the squeezing procedure to the spinful case.
In fact, one can think of several ways. Here, we will discuss the only procedure
we found to work for every model state that we considered.

Let us take an explicit example to explain our considerations, and focus on a simple 
spin singlet state, the Halperin-$(221)$ state \cite{halperin83} (using later the abbreviation $(221)$ state)  for spinful bosons, with filling
fraction $\nu=2/3$. This state is written as
\begin{equation}
\Psi_{\rm (221)} (\{ z^\ua,z^\da\}) =
\prod_{i<j} (z^\ua_i - z^\ua_j)^2 (z^\da_i - z^\da_j)^2
\prod_{k,l} (z^\ua_k-z^\da_l) \ ,
\label{halperin221}
\end{equation} 
where the $z^\ua_i$ and $z^\da_i$ denote the complex coordinates of the
$i^{\rm th}$ spin-up and -down particles, respectively. 

The $(221)$ state is the ground state of a local Hamiltonian
which can be written in terms of Haldane pseudopotentials. In particular, this
Hamiltonian projects onto states in which no two particles of the same spin
have angular momentum less than two, and no two particles of opposite spin
have relative angular momentum zero. These properties can be read off
from the wave function \eqref{halperin221}.

Let us denote by $P_{i,j}(L,S)$ the projector, which projects onto
(i.e., penalizes) the state in which particles $i$ and $j$ have relative
momentum $L$, and have overall spin $S$. In terms of
these projectors, the model Hamiltonian can be written as
\begin{equation}
H_{\rm (221)} = \sum_{i<j} P_{i,j} (0,0) + P_{i,j} (0,1) \ .
\label{h211int}
\end{equation}
The sum here is over all pairs of particles, irrespective of their spin.
We remind the reader that we are dealing with bosons, so we do not have to add
the projector $P_{i,j} (1,1)$.

For completeness, we quickly introduce the general Halperin-$(mmn)$ states, which
take the form 
\begin{equation}
\Psi_{\rm (mmn)} (\{ z^\ua,z^\da\}) =
\prod_{i<j} (z^\ua_i - z^\ua_j)^m (z^\da_i - z^\da_j)^m
\prod_{k,l} (z^\ua_k-z^\da_l)^n \ .
\label{halperinmmn}
\end{equation} 
For $m=n+1$, these states are singlet states. In general, they are the densest
zero-energy ground states of the interaction (note that the projectors now project onto $S_z$ states)
\begin{equation}
\begin{split}
H_{\rm {\rm (mmn)}} &= \sum_{i<j} \Biggl[
\sum_{0\leq p < n} P_{i,j} (p,S_z=0) + \\ &
\sum_{0\leq q < m} P_{i,j} (q,S_z=1) + P_{i,j} (q,S_z=-1)  \Biggr] \ .
\end{split}
\end{equation}

We return to the question of identifying root configurations of spinful wave functions by
considering the bosonic $(221)$ state.
Because, in this example, no orbital can be occupied by two particles,
we will use the following notation. If the $i^{\rm th}$ orbital is occupied by one spin-up particle,
we write $n_i = \ua$, and $n_i=\da$ for a down particle. An unoccupied orbital simply has
$n_i = 0$.

It has been shown that, in the thin-cylinder limit, the states that survive are those
that have electrons in neighboring sites that form singlets, separated by an empty site \cite{sy08}.
In particular, there are four configurations of the $(221)$ state of four
particles that survive in the thin torus limit, namely,
\begin{align}
\label{rp221}
&(\da,\ua,0,\da,\ua) & &(\da,\ua,0,\ua,\da)  \\
&(\ua,\da,0,\da,\ua) & &(\ua,\da,0,\ua,\da)  \ .
\nonumber
\end{align}
The partitions of the form
\begin{align}
\label{rp221forbidden}
&(\ua,\ua,0,\da,\da) & &(\da,\da,0,\ua,\ua)  
\end{align}
are absent in the $(221)$ state, because two particles of equal spin
have a minimal relative angular momentum of two, as in the bosonic Laughlin
state with $\nu=1/2$.

We will show in the next subsection how the configurations \eqref{rp221}, which 
correspond to states that survive in the thin-cylinder limit \cite{sy08},
can be used as root configurations to obtain the reduced Hilbert space.

\subsection{Squeezing rules for spinful states}

Our strategy to uniquely specify spinful states will follow the polarized case as closely
as possible, namely, we will try to find a single, or several, root partitions, from which the others
can be obtained by squeezing. On this restricted Hilbert space, we furthermore impose
the highest weight condition $L_+ \Psi = 0$. If the state is a spin singlet state, obeying $SU(2)$
invariance, we will impose the additional condition $S_+ \Psi = 0$.  As we already pointed out,
there are, in principle, several ways of doing this.
In the following, we will give a set of rules, which we found
to uniquely define a large class of model quantum Hall states, including the spin singlet Halperin
state, the non-abelian spin singlet states proposed by Ardonne and Schoutens (AS)\cite{as99,arr01},
the Haldane-Rezayi state \cite{hr88}, and a non-abelian state
exhibiting spin-charge separation \cite{all02}, which we will denote by the acronym
`SCsep'.
A lesser known fermionic spin singlet state which can be constructed
this way is the product of a permanent and a complete Jastrow factor, 
$\Psi_{\rm SFper} = {\rm Per}\bigl( \frac{1}{z_i^\ua-z_j^\da}\bigr) \times \Psi_{\rm (111)}$,
a state which was studied by Read and Rezayi \cite{rr96}. 

As examples of states which are not $SU(2)$ invariant
we mention the $(pp0)$ states  with $p>1$, and the bosonic $S_z=0$ state
$\Psi_{\rm SBper} = {\rm Per}\bigl( \frac{1}{z_i^\ua-z_j^\da}\bigr) \times \Psi_{\rm (221)}$.
Many of the states we
just mentioned turn out to have root configurations which are closely related.
We will come back to this interesting issue in section \ref{sec:rootrelations}.

We remark that, although the spin singlet composite fermion states do obey a squeezing
principle, it is not possible to uniquely define these states by imposing constraints on the
reduced Hilbert space. The reason behind this is the same as for the polarized composite
fermion states: they are not the unique ground states for any local model Hamiltonian.

We will now describe the procedure, to generate the model states, which we divide in
a few steps.
\begin{enumerate}
\item
First, one needs to decide which root configuration to use. This can simply be a choice,
or derived from a model Hamiltonian. In this root configuration, one completely
ignores the spin or internal degree of freedom. For spin-$1/2$ fermions, the maximal
occupation number in the root configuration is two, for spin-$1/2$ bosons there is
no such constraint.
\item
To construct the reduced Hilbert space, one starts by constructing all the possible states
one can obtain by squeezing from the root configuration obtained in 1. Still, one does
not take the spin degree of freedom into account (apart from the restriction in case
of fermions, as in 1.).
\item
Continue by taking all states obtained in 2., and distribute the spin degree of freedom
in all possible ways.
\item
Impose the constraints coming from the Hamiltonian, which are not taken into
account already.
\item
Impose the applicable highest weight conditions. This always includes $L_+ \Psi = 0$.
If the total spin is a good quantum number, one also needs to impose $S_+\Psi = 0$.
 \end{enumerate}

Some remarks about these steps are in order here. The procedure we employ is to
first strip off the internal degrees of freedom, perform the squeezing, and re-introduce
the internal degree of freedom. Although we seem to be working in a roundabout way,
this procedure is in fact necessary, to obtain a `large enough' reduced Hilbert space.
By this we mean that we would like our procedure to work for all known model
states with internal degrees of freedom.

As an example of a state for which the `naive' procedure does not work is the
Halperin-$(332)$ state. One of the putative root configurations reads
$(\da,0,\ua,0,0,\da,0,\ua,0,0,\da,0,\ua)$, while the other seven are obtained by
replacing  $\da,0,\ua$ with $\ua,0,\da$ in the various locations. If one starts to
squeeze the up and the down particles from these root configurations, one
never obtains a configuration like $(\ua,0,0,\ua,\da,0,0,\da,0,0,\ua,0,\da)$, which
is nevertheless present in the expansion of the $(332)$ state. Our procedure overcomes
this problem.

The fact that we first drop the internal degree of freedom, and later re-introduce them
in all possible ways, gives sometimes rise to basis-states which actually are not allowed
by the Hamiltonian. A simple example is the $(221)$ state, in which the
basis states in equation \eqref{rp221forbidden} have zero coefficient. This `problem' can
be dealt with in a simple way, by giving these basis-states which are not allowed
because of the Hamiltonian, zero coefficient by hand.
This typically only involves a low number of basis states, and only the first
few orbitals, depending on how complicated the Hamiltonian is. Typically, the number of
constraints coming from the highest weight $L_+\Psi = 0$ condition is much bigger.
In fact, explicitly setting coefficients to zero reduces the number of variables one has
to solve for. Sometimes, one does not even have to set these coefficients to zero by hand,
because these constraints are incorporated in the condition $L_+\Psi = 0$. Examples are
the $(221)$ state and the AS states.
On the other hand, for the Haldane-Rezayi and SCsep states,
one has to take additional constraints coming from the Hamiltonian into account explicitly.

The squeezing rules we presented above can be used for states without quasiholes
present, as well as states with quasiholes. The only difference lies in the root configurations
one start with. One obtains these in the same way as for polarized states with quasiholes
present.
One considers the root configuration disregarding the spin, with the appropriate number of
orbitals, and fills the orbitals such that the particles have as high an
angular momentum as possible, taking the Hamiltonian into account. This automatically gives
a configuration with the highest $L_z$ possible. The other $L_z$ sectors are obtained by
hopping the particles to lower angular momenta as explained for the polarized case at the end of section \ref{sec:polarized}. This gives a set of root configurations, all at different $L_z$. 
To obtain the reduced Hilbert spaces in the different $L_z$ sectors, one uses the same
squeezing procedure we introduced above. The number of states is then given by the
number of solutions to the constraints, namely $L_+\Psi=0$, as well as $S_+\Psi=0$
and the constraints coming from the Hamiltonian, if applicable.

\section{Explicit examples of spinful quantum Hall states}
\label{sec:singlets}

In this section, we consider a set of (spinful) states,
for which we checked that the squeezing procedure we presented in the previous
section works, and gives the right number of multiplets,
as given by the counting formulas. For singlet states, this means that
we obtain the right number of $(L,S)$ multiplets, while for states where
total spin is not a good quantum number, but $S_z$ is, we obtain the
correct number of $L$ multiplets at each possible value of $S_z$.

Underlying these counting formulas lies an exclusion \cite{h91}
(or generalized Pauli) principle,
which limits the number of particles that can occupy a certain number of
adjacent orbitals. In the polarized cases, the orbital occupations that satisfy the
exclusion principle, are precisely those orbitals that are used in the construction
of the states, using the squeezing principle. We will show that in the spinful or
multi-component case, we have exactly the same result. Namely, one can obtain the right
number of states, from an exclusion principle, but to make the correspondence work,
one needs a procedure where one first ignores the spin to generate a set of
orbital occupations. Then, one has to dress these orbital occupations with the spin
degrees of freedom, taking constraints coming from the Hamiltonian into account.
The amount of states obtained in this way, is in one-to-one correspondence to the
number of states present, for the number of particles and quasiholes under
consideration. Below, we will go
over the different states in more detail, and state in detail the constraints one
has to impose on the configurations, to obtain the correct counting. We checked
this in each case for a considerable number of particles and quasiholes, but
a proof for the claims made will be left for another occasion.

\subsection{The $(221)$ singlet states}

As we pointed out in the previous section, the bosonic
$(221)$ state is the ground state of a local Hamiltonian
that can be written in terms of Haldane pseudopotentials. 
We repeat the wave function here for convenience, and refer to the
previous section for the model interaction, eq. \eqref{h211int}
\begin{equation}
\Psi_{\rm (221)} (\{ z^\ua,z^\da\}) =
\prod_{i<j} (z^\ua_i - z^\ua_j)^2 (z^\da_i - z^\da_j)^2
\prod_{l,m} (z^\ua_l-z^\da_m) \ .
\end{equation} 

The root configuration, which one should use to generate this state, is
closely related to the configurations that survive in the Tao-Thouless
limit (see\cite{sy08} for this state).
We already discussed these configurations in the previous section,
where we described the squeezing procedure in detail. In particular, the
configurations needed are the Tao-Thouless configurations, but with the
spin degrees of freedom removed, which leads to configurations of the
form $(1,1,0,1,1,0,\ldots,0,1,1)$ in the case of the ground states (i.e., states
without additional quasiholes). The various configurations needed
for states with quasiholes are obtained in exactly the same way as the
configurations of polarized states in the presence of quasiholes, which we
explained in detail in section \ref{sec:polarized}.

The number of states generated in this way, indeed form all the ground
states of the pseudopotential Hamiltonian described above. The counting
of the number of states has been described in detail in the literature. Here,
we will formulate this counting in terms of an exclusion (or generalized Pauli)
principle \cite{h91}.

To describe this exclusion principle, which can be used to count the number
of ground states for an arbitrary number of particles and flux, we start by noting that
the filling fraction of the $(221)$ state is $\nu=2/3$.
So, we will be considering
orbital occupations in which no three neighbouring orbitals contain more than
two particles. In addition, no orbital can be occupied by two particles. By enumerating
all the configurations that satisfy these criteria, we obtain a set of configurations,
which can be grouped into a set of angular momentum multiplets. We will now turn
our attention to the question of how to  `introduce spin' to these multiplets.

We thus consider all possible ways to distribute
spin over the orbital configurations obtained from the rules above. Distributing
the spins over the orbital configurations is subjected to a constraint, namely
two neighbouring orbitals can not contain two particles with the same spin
(or better, can not form an $S=1$ multiplet), which
follows from the pseudopotential Hamiltonian.
Because the $(221)$ state is $SU(2)$ symmetric, this implies that
two particles occupying neighbouring orbitals must form a singlet pair.

The particles that are not forced to be part of a singlet pair by this rule are free,
and can be part of an arbitrary spin multiplet. To complete the counting, we thus
need to know the number of different $S$ multiplets that the free spins can form.
This is a standard problem. If one has $n$ spin-1/2 particles, the number of
$s$ multiplets is given by
\begin{equation}
\#(n,s) = \frac{2s+1}{n/2+s+1}\binom{n}{n/2+s} \ .
\label{eq:spincount}
\end{equation}
This completes the counting of the ground states of the model Hamiltonian of
the $(221)$ state, in terms of the exclusion principle outlined above.

We checked that the above is in accordance with the counting formula for the
number of (quasihole) states given the number of particles $N$ and the total
number of flux quanta $N_\phi$ on the sphere. The number of flux quanta is
given by $N_\phi = \frac{3N}{2} -2 +\frac{n}{2}$, where $n = n_\ua +n_\da$
is the total number of quasiholes, and $N = N_\ua + N_\da$.
Then, the number of states is given by
\begin{equation}
\#_{(221)} (N,n) =
\sideset{}{'}\sum_{\substack{N_\ua+N_\da=N\\n_\ua+n_\da=n}}
\binom{N_\ua+n_\ua}{N_\ua}
\binom{N_\da+n_\da}{N_\da} \ ,
\end{equation}
where the sum is over all possible ways of dividing $N$ (and $n$) into up and
down particles. In addition, the sum is constrained by the relation 
$N_\ua + n_\ua  = N_\da + n_\da$, which guarantees that both spin species see the
same amount of flux. Finally, the total $S_z$ quantum number of particular contribution
to the number of states is given by $2S_z = N_\ua - N_\da$.

It will be useful in the following to give an alternative description of the number
of spin-$s$ multiplets in the tensor product of $n$ spin-$1/2$ representations.
One of the simpler ways, out of the many ways possible, to show that 
this number is given by eq. \eqref{eq:spincount}, is as follows. The number of states with
a fixed, total value $s_z$ is given by $\tbinom{n}{(n+2s_z)/2}$. The number
of spin-$s$ multiplets is then given by the number of states with $s_z=s$ minus
the number of states with $s_z = s+1$, or  
$\tbinom{n}{n/2+s}-\tbinom{n}{n/2+s+1}= \frac{2s+1}{n/2+s+1}\tbinom{n}{n/2+s}$.

\begin{figure*}[ht]
\includegraphics[width=15cm]{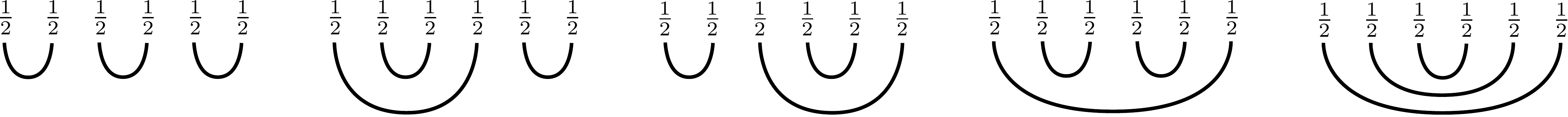}
\caption{The configurations enumerating the number of $S=0$ states in the tensor product of
$6$ spin-$1/2$ particles.}
\label{fig:s0confs}
\end{figure*}

\begin{figure*}[ht]
\includegraphics[width=15cm]{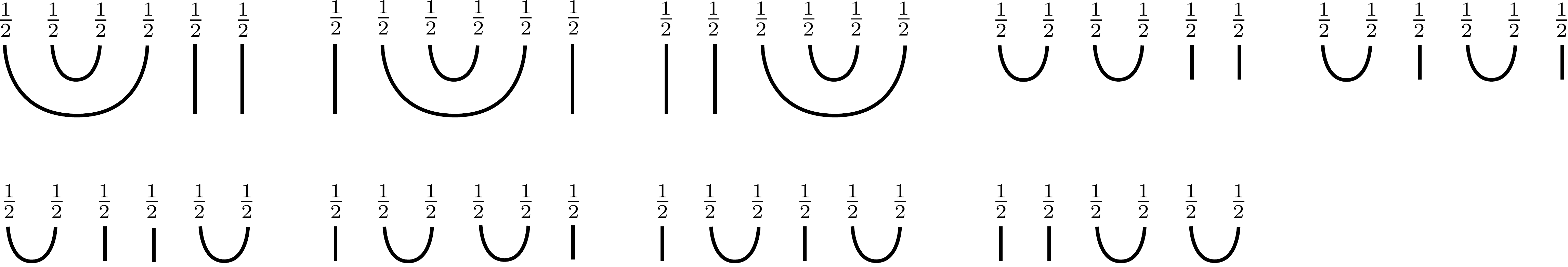}\\
\caption{The configurations enumerating the number of $S=1$ states
in the tensor product of $6$ spin-$1/2$'s}
\label{fig:s1confs}
\end{figure*}

For the non-abelian generalization of the $(221)$ state, we will need a
more graphical description of the number of spin-$s$ multiplets present in
the tensor product of $n$ spin-$1/2$ particles, which goes under the name of
the `Rumer-Pauling' rules \cite{r32,p33}.
In this representation, all the
$n$ spin-$1/2$ particles are depicted by lines, which `carry' the $SU(2)$ $s=1/2$
representation. For convenience, we order the lines next to each other. Joining
two lines, as depicted in figure \ref{fig:s0confs}, means that the two spin-$1/2$ representations
form a singlet (or valence bond).
The total number of spin singlets one can form out of $n$
spin-$1/2$ particles, is given by the number of ways one can connect the $n$
spins pairwise, such that the connecting lines do not cross. The number of such
diagrams can easily be shown to be a Catalan number, in accordance with eq.
\eqref{eq:spincount}. The total number of spin-$1$ states can be found in a similar
way, but this time, one should leave two of the spin-$1/2$ particles unpaired, and 
pair up the remaining ones%
\footnote{We remark that two parallel lines do not project onto spin-$1$.}.
Again, the lines representing the spin-$1/2$ representations
can not cross one another. In the figure \ref{fig:s1confs}
we display the diagrams enumerating the spin-$1$
diagrams. Analogously, there are five spin-$2$ configurations, and
only one spin-$3$ configuration, with all spins unpaired.


\subsection{The non-abelian spin singlet states}

One can construct non-abelian analogues of the $(221)$ spin singlet
states, in the same way as one can generalize the Laughlin $\nu=\frac{1}{2}$
state to the Moore-Read and Read-Rezayi states. The Read-Rezayi states are
labeled by a parameter $k$, which characterizes the vanishing properties of the
states, when one clusters the constituent particles. By concentrating on the simplest
bosonic state, one has that the RR-$k$ state does not vanish when $k$ particles
coincide, while the wave function vanishes quadratically when $k+1$ particles
coincide. It turns out that there is a unique, densest state with these properties.

The non-abelian spin singlet states\cite{as99} are the spin singlet analogues
of the Read-Rezayi states. The AS ground states also have the property that they do not
vanish when $k$ particles coincide (irrespective of their spin), while the wave function
vanishes quadratically (linearly) when $k+1$ particles of the same (mixed) type coincide.
An easy explicit form of the
wave function uses the `Cappelli' form \cite{cgt01} of the Read-Rezayi wave functions, which
is a symmetrized product of $k$ bosonic Laughlin $1/2$ states. Similarly \cite{sal02}, one
can write the AS states as a symmetrized product of $k$ $(221)$ states,
\begin{equation}
\begin{split}
\Psi_{\rm AS,k} (\{z^\ua,z^\da\}) &=
\mathcal{S}_{z^\ua,z^\da}
\biggl[
\Psi_{\rm (221)} (\{z^\ua_a,z^\da_a\}) \times \\ &
\Psi_{\rm (221)} (\{z^\ua_b,z^\da_b\}) \cdots
\Psi_{\rm (221)} (\{z^\ua_k,z^\da_k\}) 
\biggr] \ ,
\label{ask}
\end{split}
\end{equation}
where the $(221)$ wave function is given in eq. \eqref{halperin221} and
$\mathcal{S}_{z^\ua,z^\da}$
denotes the separate symmetrization of the spin-up particles on the one
hand, and the spin-down particles on the other.
The filling fraction of these simplest bosonic AS states is given by $\nu=\frac{2k}{3}$,
which changes to $\nu=\frac{2k}{2kM+3}$ upon multiplication of a complete Jastrow
factor for spin-up and down particles. For future
reference, we will write this factor as $\prod_{i<j} (x_i-x_j)^M$, where $x$ can denote
the position of either a spin-up or down particle.

For $k=2$, it is rather straightforward to write down an
interaction, for which the (simplest bosonic) AS states are the unique ground states. We will
concentrate on the simplest bosonic case $M=0$. For $k=2$, the interaction is a three
body interaction, which does not depend on the spin of the interacting particles, and is
identical to the model interaction having the (spinless) Moore-Read state as its ground state.
In particular, we can write
\begin{equation}
H_{{\rm AS},k=2} = \sum_{i<j<k} P_{i,j,k} (0,\frac{1}{2}) + P_{i,j,k} (0,\frac{3}{2}) \ .
\end{equation}
We do not need the term $P_{i,j,k} (1,\frac{3}{2})$, because this term will not give
a contribution to the energy because we are dealing with bosons. For arbitrary
$k$, the interactions will be a $k+1$ body interaction, penalizing the coincidence of
$k+1$ particles.  

After this short overview of the AS states, we turn our attention to the root configurations,
which survive in the Tao-Thouless limit, and which are the configurations to be used in
generating the states (with or without quasihole excitations), by using the squeezing
procedure we presented in this paper. Because the states can be written as a symmetrized
product over $k$ $(221)$ states, it naturally follows that the root configurations
(after stripping the spin degrees of freedom), can be written as $(k,k,0,k,k,0\ldots,0,k,k)$.
We have checked extensively that the number of states (or better, $(L,S)$ multiplets)
generated from the root configurations, via our procedure to construct model states
as explained in the previous section corresponds one-to-one with the counting
formula obtained from the underlying conformal field theory.
This counting formula precisely gives the
number of $(L,S)$ multiplets, given the number of particles and flux quanta. In appendix
\ref{app:counting}, we collect counting formulas for several model quantum Hall states.

The number of states can also be obtained from an exclusion principle,
analogously to the RR and Halperin states. This exclusion principle makes
use of the structure of the root configurations. As we did for the $(221)$
state, we describe the counting in the case $M=0$; multiplication of the wave functions
by an overall Jastrow factor does not change the counting, although the precise form
of the root configurations changes.

From the symmetrized expression for the AS states in eq. \eqref{ask}, 
one observes that every orbital can at most be occupied by $k$ particles,
while every set of three
consecutive orbitals can at most be occupied by $2k$ particles. These rules are
enough to determine the possible angular momentum multiplets, for a given number
of particles and number of orbitals. The more interesting part of this problem lies in
how one has to `introduce' the spin degrees of freedom to the obtained configurations.

In the Halperin states, no two up particles can occupy neighbouring orbitals, which
forces two particles occupying two neighbouring orbitals to form a singlet. In the case of
the AS states, we instead have that two neighbouring orbitals can occupy at most $k$ up
particles. This means that if two neighbouring orbitals are occupied by $k+1$ particles or
more, some of these particles will have to form singlets. Two particles forming such a
singlet have to occupy neighbouring orbitals.
This follows from the fact that the AS states are
symmetrized products of $(221)$ states, which allow, in their root configurations,
for maximally one particle per orbital. Upon symmetrization, no singlets are formed in
a single orbital. As a result, we find that some particles occupying neighbouring orbitals
are forced to form singlets.

We focus now on the remaining particles. If these particles were free to form arbitrary
multiplets, we could use eq. \eqref{eq:spincount} to obtain the number of
$S$ multiplets for each $L$ multiplet we obtained earlier. However, the `free spins',
which are not bound to form singlets, can not form arbitrary $S$ multiplets, because
we have the additional constraint that no singlet can be formed on a single site. As such,
the amount of $S$ multiplets actually depends on the precise distribution of the free
spins over the orbitals. To complete the description of the exclusion principle for the
AS states, we therefore make use of the explicit diagrams enumerating the number
of $S$ multiplets, given a number of (free) spin-$1/2$ particles, which we outlined
in the previous subsection. Given these diagrams, in which all the singlets
are completely explicit, we can simply check if they give rise to singlets on a single
site for a particular orbital occupation of the free spins. If so, the diagram does not
contribute to the number of $(L,S)$ multiplets. By making use of the rather simple
exclusion principle for the $(221)$ state, and the fact that AS states are
symmetrized products of these, we were able to obtain an exclusion principle for
the AS states. We checked the results from this method against the known counting
formula derived from the underlying conformal field theory (which also makes use
of an exclusion principle), and found complete agreement.

\section{The root configurations $(2,0^{r-1},2,0^{r-1},\ldots,0^{r-1},2)$}
\label{sec:rootrelations}

In the following subsections, we concentrate on a set
of fermionic spin singlet states, for which the root configurations are of
the form $(2,0^{r-1},2,0^{r-1},\ldots,0^{r-1},2)$, where $0^{r-1}$ denotes a
sequence of $r-1$ zeros. These states are interesting, because they
are closely related to a set of spinless (or spin-polarized), bosonic quantum
Hall states at the same filling fraction. In a recent paper\cite{hrb11}, we explained this
connection in detail for the fermionic spin singlet Haldane-Rezayi state, and
the bosonic polarized Haffnian state. Both these states can be obtained from the
root configuration $(2,0^{r-1},2,0^{r-1},\ldots,0^{r-1},2)$ with $r=4$. In the
following section, we will consider $r=3$, giving rise to the bosonic, spin
polarized `Gaffnian' wave function, while if one considers the same root configuration
for spinful fermions, one obtains a non-abelian spin singlet state, showing
spin-charge separated excitations. Finally, for $r=2$, the root configuration gives
rise to the Moore-Read state, as well as a spin singlet, fermionic permanent
state.

\subsection{The Haldane-Rezayi case}

Let us start with the Haldane-Rezayi wave function \cite{hr88}, which is a fermionic, spin singlet
$d$-wave paired state, which takes the form
\begin{equation}
\Psi_{\rm HR} (\{ z^\ua,z^\da\}) =
\Det\Bigl( \frac{1}{(z_i^\ua-z_j^\da)^2}\Bigr)
\prod_{i<j} (x_i - x_j)^2 \ ,
\label{haldane-rezayi}
\end{equation}
using the convention that the variables $x_i$ can stand for either spin-up or
down particles. The filling of the
Haldane-Rezayi (HR) wave function is $\nu=1/2$, and originally, this wave
function was proposed to describe the $\nu=5/2$ quantum Hall effect.
Nowadays, we know that this wave function describes the transition between
a gapped strong paring phase, and a weak pairing $d$-wave singlet phase\cite{rg00}.
A lot more is known about the HR wave function, which we will not dwell on here, but
instead refer the reader to the literature \cite{hr88,rg00,ww94,mr96,gfn97}.

One property we would like to point out,
is that the wave function does not vanish when a spin-up and a spin-down particle
coincide. The wave function vanishes, however, as a fourth power, when any three
particles come together (when two particles of the same spin coincide, the wave
function vanishes as a third power). 

In the following, we will focus on the connection
between the HR wave function, and the so-called Haffnian wave function, first
pointed out in \cite{hrb11}. This connection has its origin in the root configurations
needed to generate both states, as well as in the the generalized Pauli (or exclusion)
principle, which can be used to count the number of states.

Let us start by giving the interaction, for which the HR state with filling fraction $\nu=\frac{1}{2}$
is the exact ground state\cite{hr88}. The interaction assigns a non-zero-energy to any two particles
with relative angular momentum $1$. If one changes the exponent of the Jastrow factor
in eq. \eqref{haldane-rezayi} to $q$, with $q\geq 2$, the interaction that will have the
Haldane-Rezayi wave function as its unique ground state at flux
$N_\phi = qN - (q+2)$ gives non-zero-energy to any two particles with relative angular
momentum $q-1$ or $q\leq 3$. We will, however, mostly be concerned with the (fermionic) case
$q=2$. In terms of two-body projectors $P_{i,j} (L,S)$, the interaction for $q=2$ can be
written as
\begin{equation}
H_{\rm HR} = \sum_{i<j} P_{i,j}(1,0) + P_{i,j}(1,1)
\end{equation}

To generate the HR wave function via our squeezing procedure,
one has to specify the root configuration (without spin!),
which for the case at hand can be described, for $q=2$, as follows.
Each orbital is occupied by at most two particles (this follows of course from the Pauli
principle), and any sequence of four consecutive orbitals can also at most be occupied
by two particles. This leads to the following most densely packed root configuration
$(2,0,0,0,2,0,0,0,2,0,\ldots,0,2,0,0,0,2)$, corresponding to filling $\nu=\frac{1}{2}$ and
shift $\delta = 4$ (the shift being defined as $N_\phi=\nu^{-1}N - \delta$). To obtain the wave function, we use the method outlined in
the previous section. The only thing we need to specify are the additional constraints
coming from the Hamiltonian. Two particles with combined spin-$1$ can not have
relative angular momentum $1$. Indeed, from the wave function one sees that the
minimal relative angular momentum of two up (or down) particles is two. For the
squeezing rules, this implies that all configurations with $n^\ua_0 = n^\ua_1 = 1$
or $n^\da_0 = n^\da_1 = 1$ get zero coefficient. With this rule in place, we have
specified all the rules necessary to generate the zero-energy ground states of the
model interaction for the HR state, at any flux. We have verified that the amount
of zero-energy ground states corresponds exactly to the counting of such states
as performed on the sphere originally in \cite{rr96}.

To formulate an exclusion principle, which can be used to count the number of
(quasihole) states for the Haldane-Rezayi case, one has to follow the same
strategy as for the Halperin-$(221)$ and AS states. One takes the root configurations
with the spin degrees of freedom removed, and adds spin in all possible ways
consistent with the Hamiltonian. We will follow the discussion of this as given
in \cite{hrb11}. In that paper, it was shown that it does not suffice to start from the
configurations which satisfy the basic principle that each four consecutive orbitals
can be occupied by a maximum of two particles, as is the case for the root configurations
used to construct the state. In addition, one needs to consider configurations of the
form $(0,2,0,0,1)$ as well. The presence of these configurations was confirmed by
the results for the HR state on the thin-cylinder limit \cite{seidel2011}. This latter
paper also provided a counting formula for (non-localized) quasihole states on the
torus.

Following \cite{hrb11}, it was found that to formulate an exclusion principle
for the Haffnian state, it was necessary to consider these additional configurations.
They take care of the fact that the Haffnian is a so-called irrational state, with a ground
state degeneracy which grows linearly with the number of particles. For the results
on the torus, we refer to \cite{hrb11} (see also \cite{seidel2011}), and focus on the
spherical geometry here. The additional configurations can be described as
follows. Every time one has a $\nu=1/2$ Laughlin like root pattern, namely
$1,0,1,0,1,0,1,0,1$, one allows squeezing of two neighbouring particles, i.e.
$0,1,0,1,0 \rightarrow 0,0,2,0,0$, as long as one does not generate a sequence
$0,1,0,0,2$. Alternatively, one can think of the configurations $0,2,0,0,1$ as
appearing symmetrized with a $0,1,0,0,2$ configuration (but not separately
counting the latter). The basic configurations, combined with the additional ones
do account for all the ground states of the model Hamiltonian having the Haffnian
as its densest ground state. This counting was performed in \cite{thesis:green01}.

To obtain the exclusion principle for the HR state, one takes the configurations
we just described for the Haffnian, and dress them with spin in all possible ways
consistent with the model Hamiltonian. The Pauli principle implies that an orbital
occupied by two particles harbours a singlet. The Hamiltonian implies in addition
that the same is true for two neighbouring orbitals that are singly occupied, and even
for two next-nearest-neighbour orbitals that are singly occupied. Thus, for a spin
to be `free', meaning that it could be part of an arbitrary big spin-multiplet, both its
two nearest-neighbour and its two next-nearest-neighbour orbitals have to be
unoccupied. Thus, the spin of the particle occupying the middle orbital in the
configuration $1,0,0,1,0,0,1$ is free to be part of an arbitrary large spin multiplet.

The rules given above suffice to count the number of ground states of the HR
model Hamiltonian at arbitrary number of fluxes on the sphere. Namely, one
takes all the configurations allowed for the Haffnian state, and dresses them with
spin, in all possible ways consistent with the rules above. One determines which
of the spins are forced to be part of a singlet. The remaining spins form arbitrary
big spin multiplets, with a degeneracy given by, as explained in the previous
section, $\frac{2s+1}{n/2+s+1}\tbinom{n}{n/2+s}$, where $n$ is the number of
(free) spins, and $s$ the spin multiplet.

In Ref.\cite{hrb11}, it was explained that a similar reasoning indeed gives the right
ground state degeneracy on the torus, for both the Haffnian and HR states.
For the HR state, a conformal field theory description has been worked out
in \cite{ww94,mr96,gfn97}. The quasihole states can be counted by
employing the same exclusion principle.
The generalized Pauli principle we described here can also be used to
count the number of states for the Haffnian and HR state on the torus, see\cite{hrb11}.
Explicit counting formulas for these cases were given in \cite{seidel2011}.
 
\subsection{Spin-Charge separated states}
\label{sec:scsep}

By considering the root configuration $(2,0,0,2,0,0,2,0,\ldots,0,2,0,0,2)$,
which for spin-polarized bosons gives rise to the `Gaffnian' wave function\cite{src07}
one can also construct a fermionic spin singlet state. The state one obtains
in this way has been considered in the literature before, and goes under the
name of the `spin-charge separated' state, because the state exhibits
minimal quasihole excitations without spin \cite{all02}. The relevance of this
state in the realm of cold atomic gases was studied in \cite{mjr08}.
Interestingly, while the Gaffnian
state is described by a non-unitary conformal field theory, the spin-charge
separated state is obtained from a unitary conformal field theory, which is
a necessary condition for a well behaved, unitary theory describing the
edge excitations of the bulk, gapped phase.

The wave function of this state takes the form
\begin{equation}
\Psi_{\rm SCsep} (\{ z^\ua,z^\da\}) =
\pf\Bigl( \frac{1}{x_i-x_j}\Bigr) \Psi_{\rm (221)} (\{z^\ua,z^\da \}) \ ,
\label{scsep}
\end{equation}
where the Pfaffian factor is with respect to all particles. This state has filling $\nu=\frac{2}{3}$,
and the shift on the sphere is given by $3$. The interaction for which this state is
the unique, zero-energy ground state was worked out in \cite{thesis:lankvelt04}, and
can be written in terms of three body projectors\cite{ds11} $P_{i,j,k} (L,S)$, assigning energy according
to the relative angular momentum and the overall spin of the particles 
\begin{equation}
H_{\rm SCsep} =
\sum_{i<j<k} P_{i,j,k} (3,\frac{3}{2}) + P_{i,j,k} (1,\frac{1}{2}) + P_{i,j,k} (2,\frac{1}{2}) \ ,
\label{ham:scsep}
\end{equation}
where we choose to set the coefficients of the projectors to one. This Hamiltonian
penalizes the closest approach, allowed by the Pauli principle, of three up particles (say).
In addition, the two closest  approaches allowed by the Pauli principle of three particles
which form a doublet $S=1/2$ are also penalized.

To describe how we can construct this state by means of our squeezing procedure, we
have to specify the additional constraints coming from the Hamiltonian. In this case, it
turns out we have to set the coefficients of all basis states which obey
$n^\ua_0 = n^\ua_1 = n^\ua_2 = 1$ or $n^\da_0 = n^\da_1 = n^\da_2 = 1$ to zero.
In this way, we can generate all states, by squeezing from the appropriate root configuration,
which satisfy the rule that every three consecutive orbitals are occupied by at most three
particles. Solving the highest weight conditions for $L$ and $S$ gives, with the additional
constraints just given, the ground states of the Hamiltonian \eqref{ham:scsep}.

We checked that the number of states generated by our squeezing procedure indeed
gives the correct number of ground states. This counting was performed in 
\cite{thesis:lankvelt04}, the resulting counting formula will be reproduced in
appendix \ref{app:counting}. Like we did for the Haldane-Rezayi
state, we will also give an exclusion principle in this case, based on the
root configurations we employ to generate the (quasihole) states, which can also
be used to count the number of ground states of the model Hamiltonian \eqref{ham:scsep}.

In contrast to the Haldane-Rezayi case, in the case at hand, no `additional' patterns are
required to reproduce the counting. The procedure to arrive at the exclusion principle
will be equivalent to the HR case, namely we take the patterns from the related, polarized
bosonic state, and dress them with spins, taking the constraints form the Hamiltonian into
account. The related polarized bosonic state is the Gaffnian. The exclusion principle for the
Gaffnian wave function is simply that one allows all configurations, which satisfy the basic rule
that no three consecutive orbitals are occupied by three particles or more. Taking these
configurations, we assign spins in all possible ways to each configuration. Each site occupied
by two particles, will have to host a singlet. In addition, there is an additional constraint
originating in the Hamiltonian and Hilbert space constraints. In particular all configurations
with three particles of the same spin in any four consecutive orbitals are to be
discarded in the exclusion principle. This puts a constraint on configurations like
$(1,1,0,1)$ and $(1,0,1,1)$, which dictates that two of the three particles in these configurations
have to form a singlet. With these rules, one can convince oneself that one indeed
reproduces the number of ground states of the model Hamiltonian.

\subsection{Overview}

In the previous subsections, we pointed out that various states can be related to each other
via the root configurations which are used to generate these states. This gave a
relation between the non-unitary Gaffnian and a unitary spin-charge separated state, as
well as a relation between the irrational Haffnian and the non-unitary Haldane-Rezayi wave
function. Here, we will give a broader perspective by considering the root configurations
$(2,0^{r-1},2,0^{r-1},\ldots,0^{r-1},2)$, with $r$ an integer. These root configurations can be
used to generate spinless bosonic states, spin singlet fermionic states, as well as spinful
bosonic states.

\begin{table*}[ht]
\begin{tabular}{r || c | c | c | c}
& $(2,2,2)$ & $(2,0,2,0,2)$ & $(2,0,0,2,0,0,2)$ & $(2,0,0,0,2,0,0,0,2)$\\
\hline
Spinless bosons & - & ${\rm Pf}\bigl(\frac{1}{z_i-z_j}\bigr)\times \Psi_{1}$ (MR)& Gaffnian &
${\rm Hf}\bigl(\frac{1}{(z_i-z_j)^2}\bigr)\times \Psi_2$ (Haffnian)\\
$S=1/2$-fermions ($S=0$ GS) &
$\Psi_{(110)}$ & ${\rm Per} \bigl(\frac{1}{z_i^\uparrow-z_j^\downarrow} \bigr)\times \Psi_{(111)} $& 
${\rm Pf} \bigl(\frac{1}{x_i-x_j}\bigr)\times \Psi_{(221)}$ (SCsep) &
${\rm Det} \bigl(\frac{1}{(z_i^\uparrow-z_j^\downarrow)^2}\bigr)\times \Psi_{(222)} $(HR)\\
2-comp bosons ($S_z=0$ GS)& - &
$\Psi_{(220)}$ & ${\rm Per} \bigl(\frac{1}{z_i^\uparrow-z_j^\downarrow}\bigr)\times \Psi_{(221)}$ &
$\Psi_{(440)}$\\
\end{tabular}
\caption{Table with the various states one can define starting from the
$(2,0^{r-1},2,0^{r-1},\ldots,2)$ root configurations. A dash indicates that there is no $L=0$ state for a general number of particles. We remind the reader that $\Psi_{(mmn)}$ denotes the $(mmn)$ state, while $\Psi_m$ denotes the Laughlin state with filling $\nu=\frac{1}{m}$.} 
\label{tab:states}
\end{table*}

In table \ref{tab:states}, we give an overview of the states one can construct for
$r=1,2,3,4$. To generate the spinless boson states, one simply uses squeezing
to generate the reduced Hilbert space from the appropriate root configuration, and
demands that the state is an $L=0$ state. For the singlet fermionic states, one
in addition requires the states to be $S=0$ states as well. Finally, to define some
of the spinful bosonic or fermionic states, one needs to impose that some of the
states in the reduced Hilbert space have zero coefficient. We list these additional
constraints separately below:
\begin{itemize}
\item
$S=1/2$-fermions, $r=3$ (SCsep state).\\
Partitions with $n^\ua_0 = n^\ua_1 = n^\ua_2 = 1$
or with $n^\da_0 = n^\da_1 = n^\da_2 = 1$ have zero coefficient.
\item
$S=1/2$-fermions, $r=4$ (HR state).\\
Partitions with $n^\ua_0 = n^\ua_1 = 1$
or with $n^\da_0 = n^\da_1 = 1$ have zero coefficient.
\item
2-component bosons, $r=2$ (H(220) state).\\
Partitions with $n^\ua_0 = 2$ or with $n^\da_0 = 2$ have zero coefficient.
\item
2-component bosons, $r=3$ (Per$\bigl(\frac{1}{z^\ua_i-z^\da_j}\bigr)\times \Psi_{(221)}$ state).\\
Partitions with $n^\ua_0 = 2$
or with $n^\da_0 = 2$ have zero coefficient.
\item
2-component bosons, $r=4$ (H(440) state).\\
Partitions with $n^\ua_0 = 2$, $n^\da_0 = 2$,
$n^\ua_0 = n^\ua_2 = 1$ or with $n^\da_0 = n^\da_2 = 1$ have zero coefficient.
\end{itemize}

We note in passing that it is possible to construct another two-component bosonic state with $r=2$, namely
$\Psi = {\rm Pf}\bigl( \frac{1}{z_i^\ua-z_j^\ua}\bigr)
\times {\rm Pf}\bigl( \frac{1}{z_i^\da-z_j^\da}\bigr)
\times \Psi_{\rm (111)}$. This state was considered in the context of
cold atomic gases in \cite{hms11up}.
This state can be obtained from our squeezing procedure
with root partition $(2,0,2,\ldots)$, but now one needs the additional constraint that
partitions with $n^\ua_0 = n^\da_0 = 1$ have zero coefficient. This gives rise to the
state $\Psi$, but only when the number of particles is a multiple of four. If the number
of particles satisfies $N_e = 4p+2$, with $p$ an integer, then the
equations obtained from the construction above do not have a non-trivial solution,
in agreement with the fact that one can not write down the state above in this case
(at least in the absence of quasiholes). Similarly, we can construct a state at
$r=4$ of the form
$\Psi = {\rm Hf}\bigl( \frac{1}{(z_i^\ua-z_j^\ua)^2}\bigr)
\times {\rm Hf}\bigl( \frac{1}{(z_i^\da-z_j^\da)^2}\bigr)
\times \Psi_{\rm (222)}$, by squeezing from $(2,0,0,0,2,\ldots)$, and requiring that partitions with $n_0^\ua = n_0^\da = 1$, or $n_0^\ua = n_1^\da = 1$ or $n_1^\ua = n_0^\da = 1$
have zero coefficient.

In the next section, we will consider the state
$\Psi_{\rm SBper} = {\rm Per}\bigl( \frac{1}{z_i^\ua-z_j^\da}\bigr) \times \Psi_{\rm (221)}$
in some more detail. We do currently not have a Hamiltonian, for which this state is the unique zero
energy ground state. Thus, to find for instance the number of quasihole states,
we have to rely on our squeezing method to obtain these states. What we will show
in the next section is that the numbers we obtain, are in accordance with the numbers
obtained from the so-called particle entanglement spectrum calculated for the state in
the absence of quasihole excitations.

\section{Some applications: particle entanglement}
\label{sec:applications}

As one possible application of our root configuration analysis, we can compare the results that we have obtained for the quasiholes (namely, the number of quasihole states for a given number of flux quanta), with the one provided through the entanglement spectrum (ES) \cite{li2008,sterdyniak2011}.  For a single non-degenerate ground state $\ket{\Psi}$, the entanglement spectrum can be defined through the density matrix $\rho=\ket{\Psi}\bra{\Psi}$ and the decomposition of $\ket{\Psi}$ in two regions $A$, $B$. By tracing out the degrees of freedom of $B$, one obtains the reduced density matrix $\rho_A={\rm Tr}_B \rho$. Its spectrum is called the entanglement spectrum, and it unveils a rich structure of the state $\ket{\Psi}$. The key idea is to focus on one block of $\rho_A$, fixing all but one of the quantum numbers that are conserved within this operation. Then, one plots the $\xi_i$ as a function of this quantum number, where $\exp(-\xi_i)$ are the non-negative eigenvalues of $\rho_A$. Depending on the space in which the system is split into two parts, be it real, momentum, orbital or particle space, different aspects of the system excitations will be revealed through the ES.

It was shown that, if the regions $A$, $B$ are regions of particles\cite{sterdyniak2011}, the particle entanglement spectrum (PES) hence obtained by tracing over the positions of a set of $B$ particles gives information about the number of quasiholes of the system of $N_A$ particles and number of orbitals identical to that of the untraced system. In the case of the many model FQH states, the particle entanglement spectrum contains an identical number of levels as those of the quasihole
states with a reduced number of particles. This property seems to be valid even when no local Hamiltonian is known (such as the composite fermion wave functions \cite{j89}). 

\begin{figure}[t]
\includegraphics[width=\columnwidth]{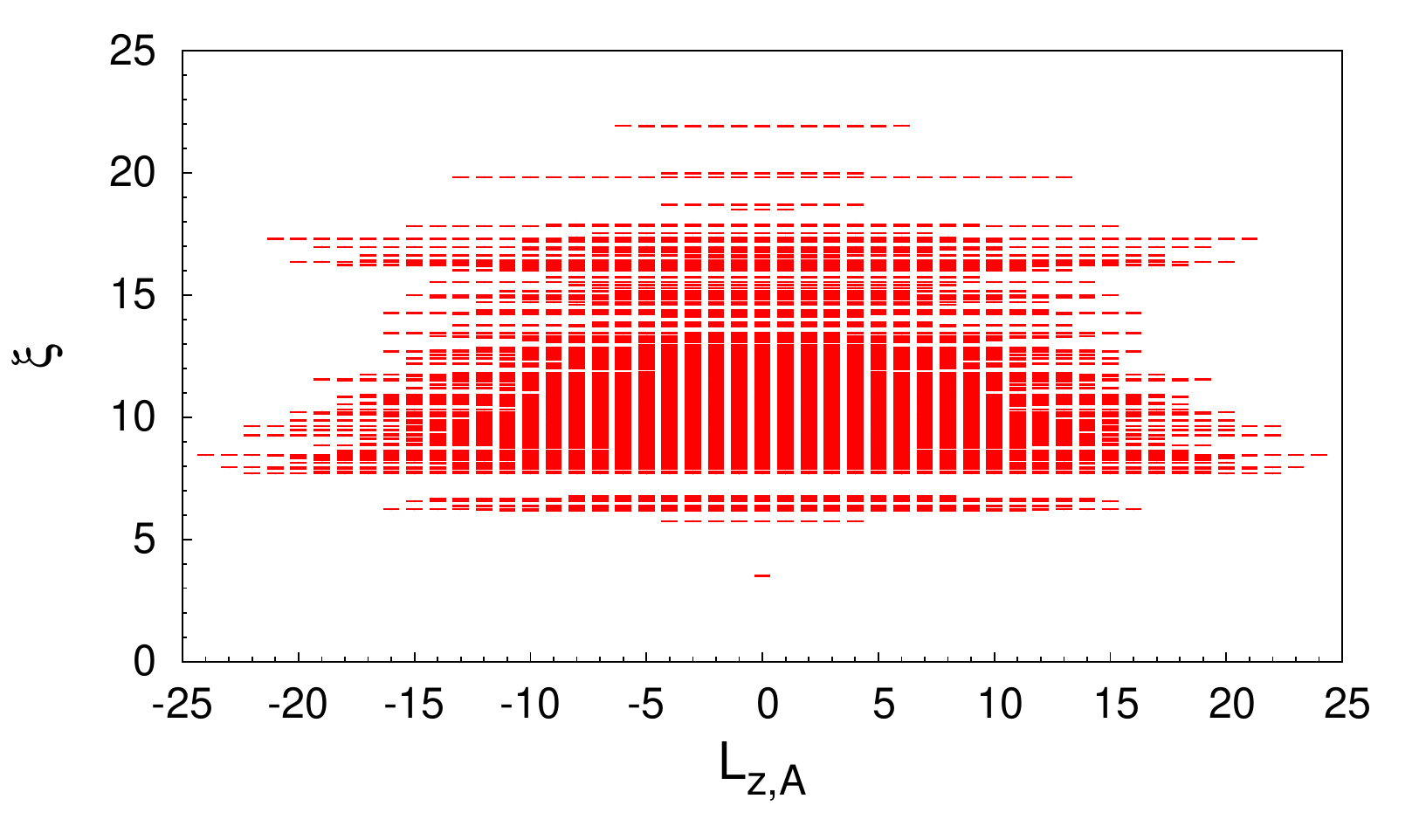}
\caption{The particle entanglement spectrum for the bosonic Haffnian state with $N=10$ particles, keeping $N_A=5$ particles. The $L_{z,A}$ degeneracy is due to the multiplet structure associated with $L^2_{A}$. The counting per value $L_{z,A}$ sector exactly matches the corresponding number of quasihole states for 5 particles and 10 added flux quanta.}
\label{fig:Haffnianspec}
\end{figure}

\begin{figure}[th]
\includegraphics[width=\columnwidth]{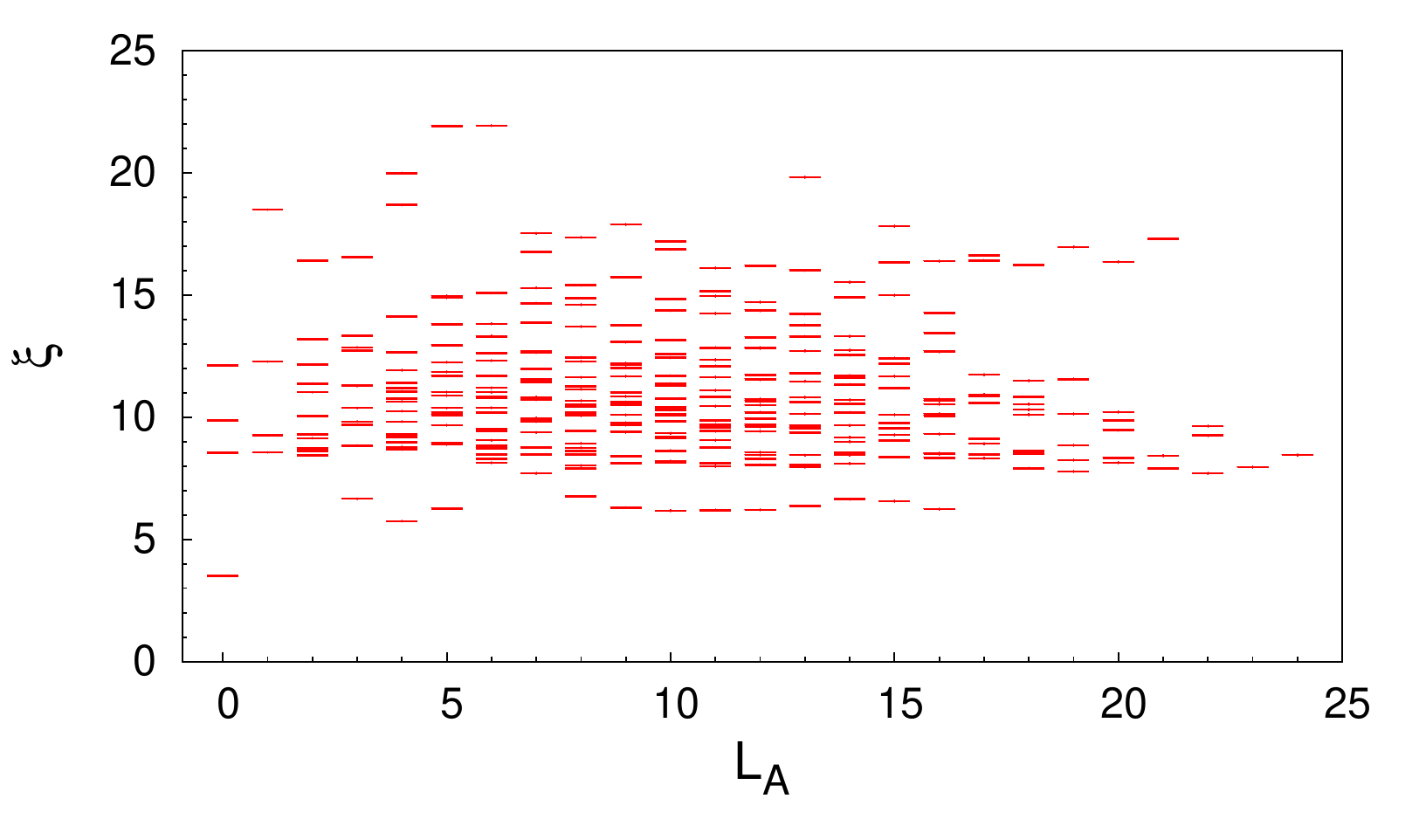}
\caption{The particle entanglement spectrum for the bosonic Haffnian state with $N=10$ particles, keeping $N_A=5$ particles. Only the entanglement levels of the highest weights are shown.}
\label{fig:Haffnianspeclzmax}
\end{figure}

In figure \ref{fig:Haffnianspec}, we show a typical ES, namely, the particle ES for the Haffnian
wave function. All the entanglement levels are plotted against the total projected angular
momentum of part $A$, $L_{z,A}$. From the figure, it is immediately clear that indeed the
total angular momentum of part $A$, $L_{A}^2$ is also a good quantum number. For comparison,
we show the same spectrum, with but with only the highest $L_z$ state of every multiplet in
figure \ref{fig:Haffnianspeclzmax}.  

It is interesting to note that the particle ES \ref{fig:Haffnianspeclzmax} shows a great deal
of resemblance to the real energy spectrum of a typical quantum Hall state on the sphere,
with a lowest lying $L=0$ state, separated by a gap from a continuum. In addition, even
a feature resembling the typical roton-mode present in the energy spectrum seems to be
present in the particle ES. The particle ES of the Haffnian state shown in figure
\ref{fig:Haffnianspeclzmax} was obtained by tracing out half of the particles.
The Haffnian can be seen as a symmetrized product of two Laughlin $\nu=\frac{1}{4}$
states. It is, thus, perhaps not so surprising that a state such as the Laughlin $\nu=\frac{1}{4}$
state should have a large contribution to the density matrix after tracing out half of the
particles. Indeed, the overlap between the state corresponding to the lowest $L=0$
entanglement level has a very large overlap with the Laughlin $\nu=\frac{1}{4}$, namely,
$\langle \Psi_{4} | \rho_{0} \rangle^2 \approx 0.999860$.
Such a feature has also been observed for the Moore-Read state \cite{sterdyniak2011}.

For the spin-polarized case on the sphere geometry, we can rely on two quantum numbers: the total angular momentum $L^2_A$ and its projection $L_{z,A}$. The additional $L^2_A$ quantum number compared to the orbital ES explains the multiplet degeneracy observed in figure \ref{fig:Haffnianspec}. The PES can be trivially extended to the spinful case. There, we have up to two additional quantum numbers that we can use, namely, the total spin $S^2_A$ if the state is a spin singlet and its projection $S_{z,A}$, which is always available. The orbital entanglement spectrum was already calculated for a spinful quantum Hall wave function, namely, the Haldane-Rezayi case \cite{thomale2010}. 

\begin{figure}[th]
\includegraphics[width=\columnwidth]{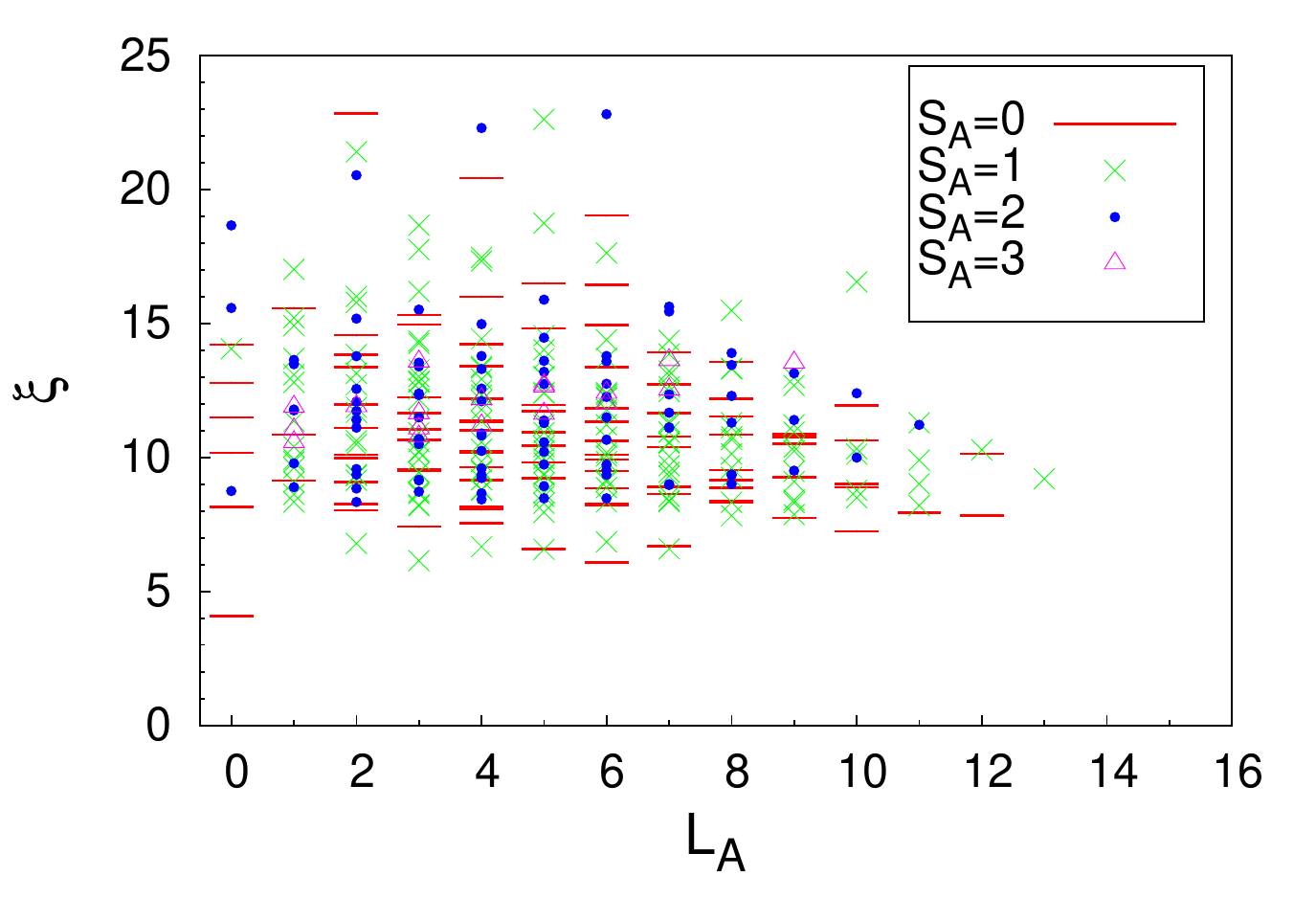}
\caption{The particle ES for the $N=12$ bosonic AS state, with $N_A = 6$.}
\label{fig:pesnass}
\end{figure}

As an example of the particle ES for a singlet state, we use the AS state, for $N=12$
particles, and trace out half of them. The spectrum is shown in figure \ref{fig:pesnass},
where we plot the highest $L_z$ and $S_z$ level for each $(L,S)$ multiplet. The lowest
entanglement level, i.e. the state contributing the most to the reduced density matrix, is
an $L=0$, $S=0$ multiplet. The $k=2$ AS state can be thought of as a symmetrized
product of two Halperin $(221)$ states. This fact is reflected in the overlap between
the $N=6$ particle $(221)$ state, and the state $|\rho_0\rangle$
corresponding to the lowest lying $L=0$, $S=0$ multiplet, which is
$\langle \Psi_{(221)}|\rho_0\rangle \approx 0.997878$.

\begin{figure}[th]
\includegraphics[width=.45\columnwidth]{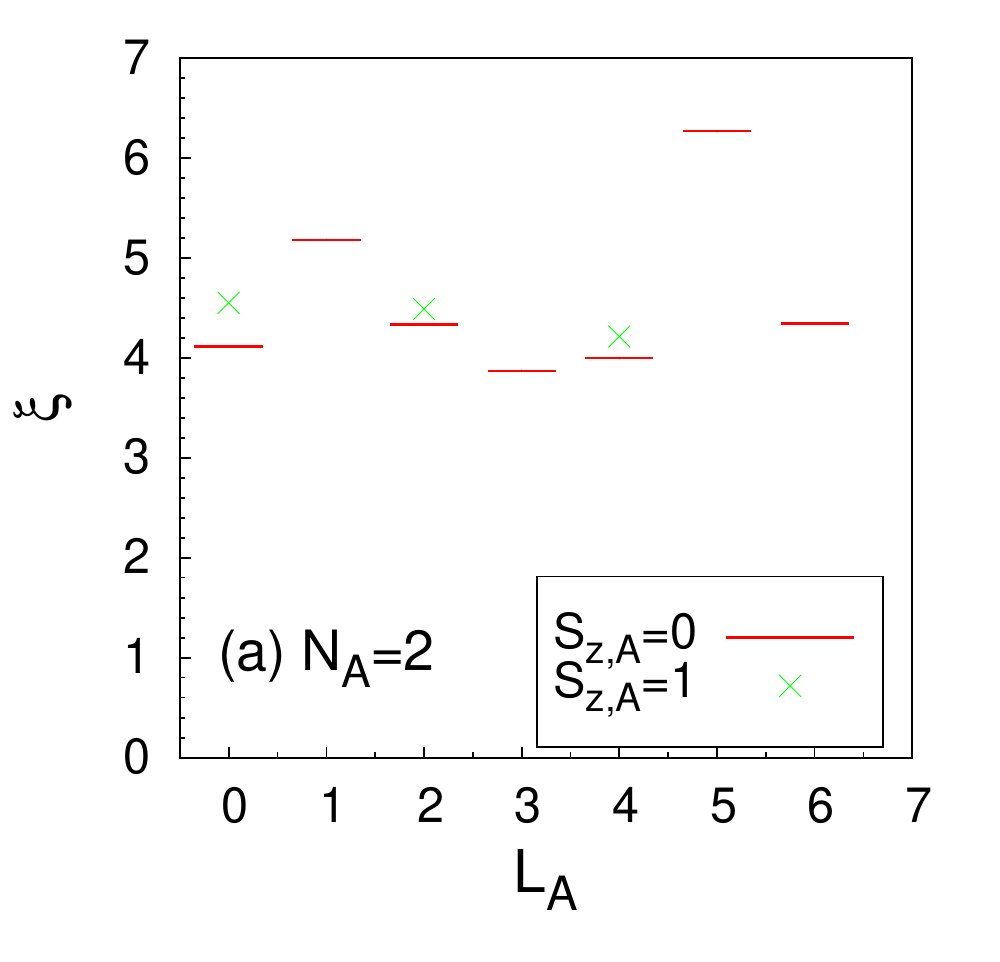}
\hspace{.05\columnwidth}
\includegraphics[width=.45\columnwidth]{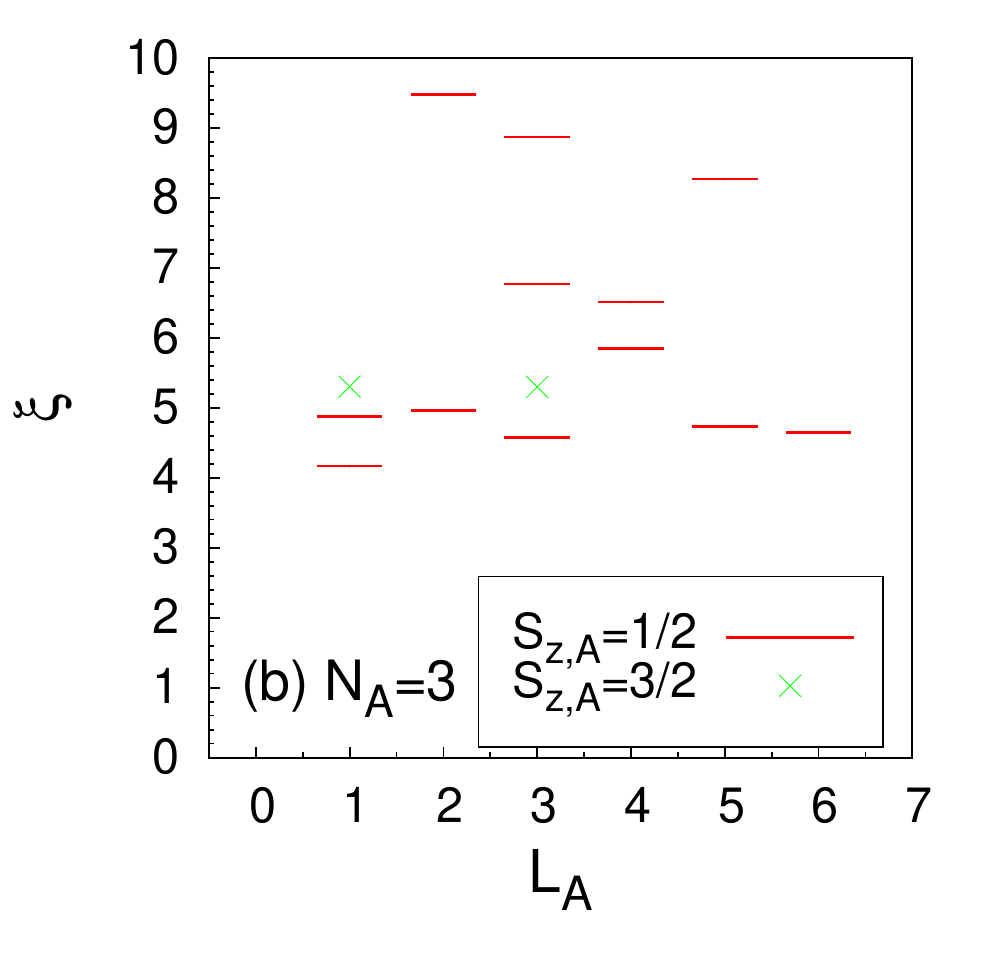}
\caption{The particle entanglement spectrum for the bosonic spin-permanent
state SBper, with $N=6$ particles, keeping $N_A=2$ (fig. a) and $N_A=3$ (fig. b) particles.}
\label{fig:spinpermspec}
\end{figure}

We will now employ the particle ES to obtain some knowledge about the spinful,
$S_z=0$ bosonic permanent state
$\Psi_{\rm SBper} =
{\rm Per} \bigl(\frac{1}{z_i^\uparrow-z_j^\downarrow}\bigr)\times \Psi_{(221)}$.
This state can be
obtained by squeezing from the root configuration $(2,0,0,2,0,\ldots,0,2,0,0,2)$. Then,
requiring a state to be an $L=0$ state, and that no two particles with the same spin
have relative angular momentum smaller than two (and thus that configurations with
$n^\ua_0 = 2$ or $n^\da_0 = 2$ have coefficient zero), leads to a unique state,
the $S_z=0$ bosonic permanent state. We checked this statement for small particle
numbers. One way to analyze this state, would be to find a model Hamiltonian for which
this state is the unique ground state. With this model Hamiltonian, one can check the
number of quasihole states, upon adding flux in comparison to the state without quasiholes.
These numbers can then be compared to the number of states one obtains from the
squeezing procedure we presented in this paper. Another way of comparing the number
of quasihole states is to make use of the connection between the level counting of the
particle ES, and the number of quasihole states, which has been shown to hold for all
model states so far. To this end, we calculated the particle entanglement spectrum for the
state
$\Psi_{\rm SBper} = {\rm Per} \bigl(\frac{1}{z_i^\uparrow-z_j^\downarrow}\bigr)\times \Psi_{(221)}$
with six particles.
Figure \ref{fig:spinpermspec} shows the particle ES for system $A$ consisting of
two and three particles in parts (a) and (b) respectively. 

\begin{figure}[th]
\includegraphics[width=\columnwidth]{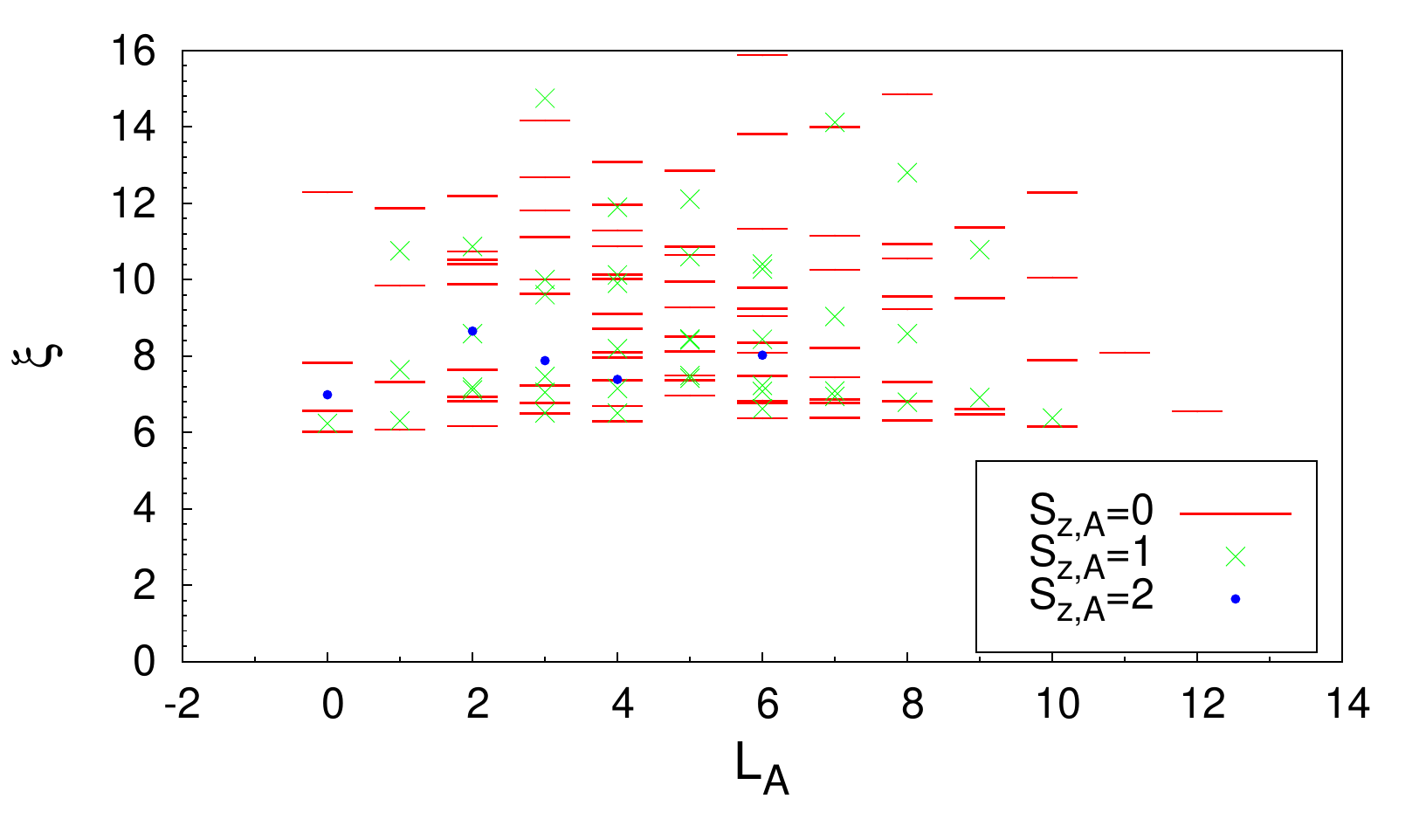}
\caption{The particle entanglement spectrum for the bosonic spin-permanent
state SBper, with $N=8$ particles, keeping $N_A=4$.}
\label{fig:spinpermspec8}
\end{figure}

In these particle entanglement spectra, we only plot the maximum $L_{z,A}$ state of
each multiplet for clarity. The red lines indicate $S_{z,A} = 0$ ($S_{z,A}=1/2$) states,
the green crosses $S_{z,A} = 1$ ($S_{z,A}=3/2$) state, for $N_A=2$ ($N_A=3$).
This state is not a spin singlet state, so $S^2_{A}$ is not a good quantum number.
The number of $(L,S_z)$ multiplets for the two cases are given in table \ref{tab:spinper}.
In figure \ref{fig:spinpermspec8}, we show the particle ES in the case of eight particles, and
$N_A=4$.

\begin{table}
\begin{tabular}{r|rcccccc}
$N_a = 2$ & $L=0$ & $1$ & $2$ & $3$ & $4$ & $5$ & $6$ \\
\hline
$S_{z,A}= 0$ & 1 & 1 & 1 & 1 & 1 & 1 & 1\\
$1$ & 1 & 0 & 1 & 0 & 1 & 0 & 0\\
\hline
\hline
$N_a = 3$ & $L=0$ & $1$ & $2$ & $3$ & $4$ & $5$ & $6$ \\
\hline
$S_{z,A}= 1/2$ & 0 & 2 & 2 & 3 & 2 & 2 & 1\\
$3/2$ & 0 & 1 & 0 & 1 & 0 & 0 & 0\\
\end{tabular}
\caption{Number of particle ES $(L,S_z)$ multiplets for the bosonic spin-permanent state with
$N_{A}=2$ (top) and $N_{A}=3$ (bottom). }.
\label{tab:spinper}
\end{table}

The total number of flux quanta for the bosonic spin-permanent state is
$N_\phi= \frac{3N_{e}}{2}-3$. For six particles, $N_\phi = 6$. So, if we want
to compare the number of levels in the particle ES, we should compare with the
number of states obtained from the squeezing procedure, with $2$ and $3$ particles,
for $N_\phi = 6$. The root configurations one should use as a starting point for the
squeezing procedure are
\begin{equation*}
\begin{tabular}{r|cc}
& $N=2$ & $N=3$\\
\hline
$L_z = 6$ & (0,0,0,0,0,0,2) & (0,0,0,1,0,0,2) \\
$L_z = 5$ & (0,0,0,0,0,1,1) & (0,0,1,0,0,0,2) \\
$L_z = 4$ & (0,0,0,0,1,0,1) & (0,1,0,0,0,0,2) \\
$L_z = 3$ & (0,0,0,1,0,0,1) & (1,0,0,0,0,0,2) \\
$L_z = 2$ & (0,0,1,0,0,0,1) & (1,0,0,0,0,1,1) \\
$L_z = 1$ & (0,1,0,0,0,0,1) & (1,0,0,0,1,0,1) \\
$L_z = 0$ & (1,0,0,0,0,0,1) & (1,0,0,1,0,0,1) 
\end{tabular}
\end{equation*}
By performing the procedure we outlined above, we obtained a number of states
for each possible value of $L$ and $S_z$, which is in complete accordance to the
number of multiplets obtained from the particle entanglement spectrum. We checked this
for both $N_{e} = 6$ and $N_{e}=8$, which gives us a non-trivial consistency check on
the squeezing procedure we proposed, where
we used a state for which (at the moment) no other approaches such as a
conformal field theory approach, is available. It seems likely, however, that a conformal
field theory description is possible. Most likely, such a description would rely on a non-unitary
conformal field theory, which could serve as a check on the results obtained above.

\section{Conclusion and Outlook}

In this article, we have generalized the concept of root partitions and squeezing, known for spinless states, to the case of spinful quantum Hall states. We have checked for several model states that our procedure leads to the right wave function both for the ground state and the quasihole states. In particular, we have stressed that the naive generalization, i.e., keeping the spin information during the squeezing procedure, may fail. Thus, one has to rely on an undressed root partition, proceed with the squeezing, and then dress the configurations with spin, in a way that is compatible with the Hamiltonian.

We have looked at several model states, such as the Halperin states and non-abelian spin singlet states, to test the validity of our set of rules. Using these spinful root partitions, we have provided a spin-$1/2$ generalization of the spinless $(k=2,r)$ sequence, which includes the Moore-Read state, Gaffnian and Haffnian states. As an application, we have shown that the counting observed when performing the particle entanglement spectrum on the ground state exactly matches the counting of the quasihole states relying on our rules. In addition, this counting also matches the counting
results obtained by counting the number of zero-energy states of the model Hamiltonian
for the state under consideration, in the cases when such a Hamiltonian is available.

We hope that our method will provide a way to study topological phases with internal degrees
of freedom, and shed light on some poorly understood quantum Hall wave functions, such
as the irrational Haffnian wave function, via the connection with better understood wave
functions such as the non-unitary Haldane-Rezayi wave function. In addition, it would be
interesting to compare our method in detail with other methods (inspired by the
question of classifying the possible topological phases), such as the
`pattern of zeros' approach \cite{bw08a,bw08b,bw10,bw10up} (see, also, \cite{s10})
and generalization of the Jack polynomials\cite{be11}.
Another interesting question is the generalization of the series $(k=2,r)$ for $r>4$,
since the polarized case already displays a rich structure such as its connection to
the $N=1$ superconformal theories\cite{ers10} for $r=6$.

{\em Acknowledgements.} We thank S. Simon for discussions, and for stimulating
one of us (EA) to make the mathematica packages available. NR acknowledges B.A. Bernevig and B. Estienne for 
fruitful discussions. EA thanks the Laboratoire Pierre Aigrain, ENS for hospitality.

\appendix
\section{Spin-singlet states}
\label{app:raising-lowering}

We will start by a brief description of spin singlet states. In the following, the coordinates
$z^\ua_i$ will
denote the spin-up particles, while the $z^\da_j$ denote the spin-down particles.
The wave functions
are composed of the orbital part, $\Psi (\{z^\ua_i,z^\da_j\})$ which is a polynomial in
the $z^\ua_i$'s and $z^\da_j$'s,
as well as a spin part, which we usually omit. The spin part has the first $N_\ua$ spins up, and the
following $N_\da$ spins down. We will omit the usual exponential factors.  

We will state the condition on $\Psi (\{z^\ua_i,z^\da_j\})$,
in order for the state to be a spin singlet. We
assume that we are dealing with either bosons or fermions states. We will be concerned with the
symmetry properties under the exchange of $z^\ua$'s with $z^\da$'s.
For the state to be a singlet, acting with both spin-raising and -lowering
operators should give zero. Acting with the spin-raising operator has the following effect on the
orbital part $\Psi (\{z^\ua_i,z^\da_j\})$ of the wave function.
A spin-down particle, say $z^{\da}_{N_\da}$ has to be raised to 
become a spin-up particle, which means it has to be symmetrized (anti-symmetrized)
with all spin-up particles:
\begin{equation}
\label{sraise}
S^{+} \Psi (\{z^\ua_i,z^\da_j\}) =
\Psi (\{z^\ua_i,z^\da_j\}) \pm \sum_{i=1}^{N_\ua} \Psi(z^\ua_i \leftrightarrow z^\da_{N_\da}) \ ,
\end{equation}
where in the bosonic (fermionic) case, one needs the plus (minus) sign.
We will implicitly assume that the variable $z^\da_{N_\da}$ will be renamed to $z^\ua_{N_\ua+1}$,
to incorporate the effect that the number of spin-up (-down) particles was increased (decreased)
by one. Similarly, we have the spin-lowering operator
\begin{equation}
\label{slower}
S^{-} \Psi (\{z^\ua_i,z^\da_j\}) =
\Psi (\{z^\ua_i,z^\da_j\}) \pm \sum_{j=1}^{N_\da} \Psi(z^\ua_{N_\ua} \leftrightarrow z^\da_{j}) \ ,
\end{equation}
where we assume that $z^\ua_{N_\ua}$ is renamed to $z^\da_{N_\da+1}$.
The condition for the state to be a spin singlet is now easily written
down. First, to have $S_z=0$, we need to have
$N_\ua=N_\da$. Secondly, both spin-raising and -lowering operators should give zero:
\begin{align}
\label{fockcyclic}
S^{+} \Psi (\{z^\ua_i,z^\da_j\}) & = 0 &
S^{-} \Psi (\{z^\ua_i,z^\da_j\}) & = 0 \ .
\end{align}
The conditions \eqref{fockcyclic} go under the name of the Fock-cyclic conditions, and were spelled
out in detail in ref. \onlinecite{lm62}. Note that in this paper, we will not be concerned with the Young
`symmetrization' procedure. In the case of quasihole states, we will have to consider multiplets of
both spin and angular momentum. In that case, to obtain the highest spin-state, we only need to
consider the action of the spin-raising operator. This is actually also true for in the spin singlet case,
because the polynomials we will consider will be (anti) symmetric under exchange of all
spin-up particles with all the spin-down particles.

For completeness, we recall that angular momentum raising and lowering operators
(on the sphere) take the following form
\begin{align}
& L^{-} \Psi (\{z^\ua_i,z^\da_j\}) =
\biggl( \sum_{i=1}^{N_\ua} \partial_{z^\ua_i} + \sum_{j=1}^{N_\da} \partial_{z^\da_j} \biggr)
\Psi (\{z^\ua_i,z^\da_j\}) \\
& L_z \Psi (\{z^\ua_i,z^\da_j\})  = \\ \nonumber &
\biggl(
N N_\phi/2 - \sum_{i=1}^{N_\ua} z^\ua_i \partial_{z^\ua_i} - 
\sum_{j=1}^{N_\da} z^\da_j \partial_{z^\da_j}
\biggr)
\Psi (\{z^\ua_i,z^\da_j\}) \\
&L^{+} \Psi (\{z^\ua_i,z^\da_j\}) =  \biggl(
N_\phi \sum_{i=1}^{N_\ua} z^\ua_i + N_\phi \sum_{j=1}^{N_\da} z^\da_i - \\ & \nonumber
\sum_{i=1}^{N_\ua} (z^\ua_i)^2 \partial_{z^\ua_i} 
- \sum_{j=1}^{N_\da} (z^\da_j)^2 \partial_{z^\da_j}
\biggr)  \Psi (\{z^\ua_i,z^\da_j\}) \ ,
\end{align}
where $N_\phi$ is the number of flux quanta, and the total number of particles is given
by $N=N_\ua+N_\da$.

\section{Numerical implementation}
\label{app:numerics}

Having the explicit form of the raising and lowering operators, see appendix
\ref{app:raising-lowering}, it is now a straightforward
matter to implement the squeezing procedure we introduced in this paper numerically.
From the form of raising and lowering operators $L^+$ and $L^-$, it is clear that it is
easiest to implement $L^-$, and demand that the states are lowest weight states, which
is of course completely equivalent with demanding states to be highest weight.

In practice, one has to implement the form of $L^-$ and $S^+$ on arbitrary symmetric
or anti-symmetric monomials, depending on whether one is considering bosons or
fermions. We have implemented these routines, as well as some others, in a mathematica
package, which is available for download \cite{packages}. These routines include
solving routines, which find the solutions for the highest weight constraints.

\section{A collection of counting formulas}
\label{app:counting}

In this appendix, we will collect, for convenience, the counting formula's for the
number of states of the various model quantum Hall states we considered in this
paper. After introducing some notation, we will start with some polarized states,
in particular the Read-Rezayi states for
arbitrary $k$ (including the Laughlin and Moore-Read cases), followed by the
characters for the (polarized states obtained from the root configurations
$(2,0^{r-1},2,\ldots,0^{r-1},2)$, for $r=2,3,4$, i.e the Moore-Read, Gaffnian and
Haffnian states.

We will continue with some spin singlet states, first the AS states for arbitrary $k$
(including for $k=1$ the Halperin-$(221)$ state), again followed by states
obtained from the root configurations $(2,0^{r-1},2,\ldots,0^{r-1},2)$, in this case
the fermionic spin singlet states, i.e. the permanent state for $r=2$, the spin-charge
separated state for $r=3$, and the HR state for $r=4$

\subsection{Some notation}

The character formulas are stated in terms of `q-binomials', which are
q-deformations of the ordinary binomials, and keep track of the $L_z$
angular momentum. We will first introduce the notation $(q)_m$, for $m$
a positive integer, $(q)_m = \prod_{i=1}^{m}  (1-q^i)$. In addition, we define
$(q)_0=1$. The q-binomial is defined as
\begin{equation}
\begin{bmatrix} a \\ b \end{bmatrix} =
\begin{cases}
\frac{(q)_a}{(q)_{a-b}(q)_{b}} & \text{for $a,b\in \mathbb{N}$ and $0\leq b \leq a$}, \\
0 & \text{otherwise}.
\end{cases}
\end{equation}

For instance, the number of states with $f$ fermions in $N_\phi+1$ orbitals is given
by $\tbinom{N_\phi+1}{f}$.
Assigning the $l_z$ angular momenta $-N_\phi/2,N_\phi/2+1,\ldots,N_\phi/2$
to the orbitals, as is applicable for quantum Hall states on the sphere with $N_\phi$
flux quanta, one finds that the number of states is generated by
$q^{-(N_\phi+1-f)f/2} \begin{bmatrix} N_\phi+1 \\ f\end{bmatrix}$.
Namely, this expression
can be expanded as $\sum_{l=-(N_\phi+1-f)f/2}^{(N_\phi+1-f)f/2} c_l \, q^l$,
where $l$ runs over half-integers if both $N_\phi$ and $f$ are odd. Otherwise,
$l$ runs over the integers. The numbers $c_l$ are equal to the number of states
with $L_z = l$. In addition, these states can be organized into angular momentum
multiplets, because, for $l\geq 0$, one has $c_l \geq c_{l+1}$, and $c_{l} = c_{-l}$.
As an example, the number of states for
$2$ fermions in $6$ orbitals is given by
$q^{-4} \begin{bmatrix} 6 \\2 \end{bmatrix} = q^{-4} + q^{-3} + 2q^{-2}+ 2q^{-1}+3+
2q+ 2q^2 + q^3+q^4$. This gives rise to one $L=4$, one $L=2$ and one $L=0$ multiplet.

The number of states of $b$ bosons with $N_\phi$ flux, or in $N_\phi+1$ orbitals, is
similarly given by $q^{-N_\phi b/2} \begin{bmatrix}N_\phi+b\\ b \end{bmatrix}$.

Finally, we will make use of the following notation in the subsequent sections.
The matrix $\cM_{k}$ has dimensions $k\times k$, and elements $(\cM_{k})_{i,j} = \min(i,j)$.
The matrix $\cO_{k}$ has dimensions $k\times k$, and elements $(\cO_{k})_{i,j} = \max(0,i+j-k)$.

\subsection{The Read-Rezayi state}

We will start out with the character for the Read-Rezayi states, with parameter $k$.
The `basic' bosonic RR states, i.e., those without any overall Jastrow factor, have
filling fraction $\nu=\frac{k}{2}$. The number of flux quanta for these states is given
by $N_\phi = \frac{2}{k} N -2 + \frac{n}{k}$, with $N$ the number of particles and
$n$ the number of quasiholes.
We note that for $N$ not a multiple of $k$, $n$ has to be non-zero, in order that the
number of flux quanta is an integer. We note that,
the counting of quasihole states remains unchanged if the wave function is multiplied
by an overall Jastrow factor. We therefore write the formulas in terms of $N$ and $n$,
the number of quasiholes, and not in terms of the number of flux quanta, because
the latter will change upon multiplying the wave function by an overall Jastrow factor.

The counting formula for the number of (quasihole) state in the RR case is given by\cite{a02}
\begin{equation}
\begin{split}
\#_{\rm RR} (N,n,k) &= q^{-\frac{(2N+n)N}{2k}}
\sum_{\substack{a_1,\ldots,a_k \geq 0\\\sum_{i=1}^{k} i a_i=N}}
q^{\ba \cdot \cM_k \cdot \ba} \times \\ & 
\prod_{j=1}^{k}
\begin{bmatrix}
j \frac{2N+n}{k} -2 (\cM_{k}\cdot \ba)_j + a_j \\ a_j
\end{bmatrix}
\end{split}
\label{eq:rrk-counting}
\end{equation}
Here, the vector $\ba$ is given by $\ba = (a_1,a_2,\ldots,a_k)$.

\subsection{The root configurations $(2,0^{r-1},2,0^{r-1},\ldots,0^{r-1},2)$: polarized bosonic states} 
\label{sec:app-pb}

For $r=2$, this case equals the Moore-Read cases, which in turn can be though of
as the $RR$ state for $k=2$. Here, we will display the ($q$-deformed version of the)
form of the counting formula as
it originally appeared in \onlinecite{rr96}. The formula \eqref{eq:rrk-counting}
in the previous section with $k=2$
yields a different, but equivalent expression
\begin{equation}
\#_{\rm MR} (N,n) = q^{-\frac{Nn}{4}}
\sum_{f} 
q^{\frac{1}{2}f^2}
\begin{bmatrix}
\frac{n}{2} \\ f
\end{bmatrix}
\begin{bmatrix}
\frac{N-f}{2}+n \\ n
\end{bmatrix}
\end{equation}
The sum over $f$ runs over even (odd) integers for $N$ even (odd), and the number of
quasiholes $n$ is always even.

For $r=3$, we obtain the Gaffnian wave function, for which the
the number of flux quanta is given by $N_\phi = \frac{3N}{2}-3+n$.
The counting formula was derived in \cite{src07} and 
we will display its q-deformed version here, which is valid for $n>0$.
\begin{equation}
\begin{split}
\#_{\rm Gaffnian} (N,n) &= q^{-\frac{Nn}{4}}
\sum_{f} 
q^{\frac{f}{2}(\frac{f}{2}+1)} \times \\ &
\begin{bmatrix}
\frac{n+f}{2}-1 \\ f
\end{bmatrix}
\begin{bmatrix}
\frac{N-f}{2}+n \\ n
\end{bmatrix}
\end{split}
\end{equation}
For $n=0$, there is only one state for $N$ even, and no states otherwise.
This case can be included in the formula, if we define $\begin{bmatrix}a \\ 0 \end{bmatrix} = 1$
for all integers $a$, even when $a<0$. For $N$ odd, the minimal
number of quasiholes required to have a state is three. We note that $N$ and $n$ have the
same parity.

The case $r=4$ corresponds to the Haffnian wave function, which was considered
in detail in \cite{thesis:green01}, where the counting was performed.
The number of flux quanta is given by $N_\phi = 2 N -4 + \frac{n}{2}$.
The counting formula is given by
\begin{equation}
\#_{\rm Haffnian} (N,n) = q^{-\frac{Nn}{4}}
\sum_{b} 
q^{b}
\begin{bmatrix}
b+\frac{n}{2}-2 \\ b
\end{bmatrix}
\begin{bmatrix}
\frac{N-b}{2}+n \\ n
\end{bmatrix}
\end{equation}
In order for this formula to be valid in all cases, we again have to define 
$\begin{bmatrix}a \\ 0 \end{bmatrix} = 1$
for all integers $a$. For $N$ odd, $n$ has to be at least four.

\subsection{The non-abelian spin singlet states}
We will continue with a set of spin singlet states, which analogously to the Read-Rezayi
states, can be defined for an arbitrary integer $k$. For $k=1$, they reduce to the
Halperin-$(221)$ states. The filling fraction of these states is $\nu = \frac{2k}{3}$
(in their simplest bosonic version). The flux is given by
$N_\phi = \frac{3}{2k}N-2+\frac{n}{2k}$, with $N$ the total number of particles,
$N = N_\ua + N_\da$, and $n$ the total number of quasiholes, $n=n_\ua+n_\da$.
There is a constraint on these numbers, namely $N_\ua+n_\ua = N_\da+n_\da$,
which implies that the flux seen by the spin-up particles is the same as the flux
seen by the spin-down particles. The counting formula for the number of states
is given by\cite{a02}
\begin{equation}
\begin{split}
&\#_{\rm AS} (N,n,k) = \\
& q^{-\frac{(3N+n)N}{4k}}
\sideset{}{'}\sum_{\substack{
N_\ua + N_\da = N\\
n_\ua+n_\da = n\\
a_1,\ldots,a_k \geq 0\\
b_1,\ldots,b_k \geq 0\\
}}
s^{\frac{N_\ua-N_\da}{2}}
q^{\ba \cdot \cM_k \ba+\bb \cdot \cM_k \bb - \ba \cdot \cO_k \bb}\times \\&
\prod_{j=1}^{k}
\begin{bmatrix}
j\frac{2N_\ua +N_\da+n_\ua}{k} - (2\cM_k\cdot \ba+\cO_k\cdot \bb)_j+ a_i \\ a_i
\end{bmatrix}
\times \\&
\begin{bmatrix}
j\frac{N_\ua +2N_\da+n_\da}{k} - (2\cM_k\cdot \bb+\cO_k\cdot \ba)_j+ b_i \\ b_i
\end{bmatrix}
\end{split}
\end{equation}
where the prime denotes the constraints
$\sum_{i=1}^{k} i a_i=N_\ua$, $\sum_{i=1}^{k} i b_i=N_\da$ and
$N_\ua+n_\ua=N_\da+n_\da$. The vectors $\ba$ and $\bb$ are given by
$\ba = (a_1,\ldots,a_k)$ and $\bb = (b_1,\ldots,b_k)$.
The exponent of $s$ gives the $S_z$ quantum number of the particular contribution
to the number of states. Having access to both the $L_z$ and $S_z$ quantum numbers,
one can extract the number of $(L,S)$ multiplets, present for arbitrary number of
flux quanta.

\subsection{The root configurations $(2,0^{r-1},2,0^{r-1},\ldots,0^{r-1},2)$: $S=0$ fermionic states} 

Like in section \ref{sec:app-pb}, we will define
$\begin{bmatrix} a \\ 0 \end{bmatrix} = 1$ for all integers $a$.

We will start with the case $r=2$, which corresponds to the
$\nu=1$ fermionic singlet permanent state
$\Psi_{\rm SFper} ={\rm Per} \bigl(\frac{1}{z_i^\uparrow-z_j^\downarrow} \bigr)\times \Psi_{(111)}$ .
The number of flux quanta is given by $N_\phi = N - 2 + \frac{n}{2}$.
The model Hamiltonian as well as the counting formula were given in
\cite{rr96}. Here, we give the q-deformed version in a slightly different form
\begin{equation}
\begin{split}
\#_{\rm SFper} (N,n) & = q^{-\frac{Nn}{4}}
\sum_{b_\ua,b_\da \geq 0}
s^{\frac{b_\ua-b_\da}{2}}
q^{\frac{b_\ua+b_\da}{2}} \times \\ &
\begin{bmatrix}
b_\ua +\frac{n}{2} -1 \\ b_\ua
\end{bmatrix}
\begin{bmatrix}
b_\da +\frac{n}{2} -1 \\ b_\da
\end{bmatrix}
\begin{bmatrix}
\frac{N-b_\ua-b_\da}{2} + n \\ n
\end{bmatrix}
\end{split}
\end{equation}
The structure resembles the structure of the counting formula for the Haffnian.
In particular, it is expected that the number of states without quasiholes on the
sphere, grows linearly with the number of particles, indicating that this state
is also irrational. 

The case $r=3$ corresponds to the (unitary) spin-charge separated state
of \cite{all02}. The number of flux quanta for this state is given by
$N_\phi = \frac{3}{2}N - 3 + \frac{n_\ua+n_\da+n_h}{2}$, where $N = N_\ua + N_\da$
the number of particles, while $n_\ua$, $n_\da$ and $n_h$ are the number of
up, down, and charged but spinless quasiholes. The total number of quasiholes
$n=n_\ua + n_\da + n_h$ has the same parity as $N$.
The counting was worked out in \cite{thesis:lankvelt04}, with
the following result,
\begin{equation}
\begin{split}
& \#_{\rm SCsep}(N,n) = q^{-\frac{N n}{4}}
\sideset{}{'}
\sum_{
\substack{
N_\ua + N_\da = N \\
n_\ua + n_\da + n_h = n\\
f\geq 0}}
s^{\frac{N_\ua-N_\da}{2}}
q^{\frac{f^2}{2}+\frac{(n_\ua+n_\da)^2}{4}} \times \\ &
\begin{bmatrix}
\frac{n_h}{2} \\ f
\end{bmatrix}
\begin{bmatrix}
N_\ua-n_\da + n_\ua \\ n_\ua
\end{bmatrix}
\begin{bmatrix}
N_\da-n_\ua + n_\da \\ n_\da
\end{bmatrix}
\begin{bmatrix}
\frac{N-f}{2}+n_h \\n_h
\end{bmatrix} \ ,
\end{split}
\end{equation} 
where the prime indicates the constraint $N_\ua + n_\ua = N_\da + n_\da$.

Finally, we come to $r=4$, namely the Haldane-Rezayi case. The counting for this
state was worked out in \cite{rr96}. The number of flux quanta is given by
$N_\phi = 2 N -4 + \frac{n}{2}$, with $n$ the number of
quasiholes. The counting formula reads

\begin{equation}
\begin{split}
\#_{\rm HR} (N,n) &= q^{-\frac{Nn}{4}}
\sum_{f_\ua,f_\da \geq 0} 
s^{\frac{f_\ua-f_\da}{2}}
q^{\frac{f_\ua^2+f_\da^2+f_\ua+f_\da}{2}} \times \\ &
\begin{bmatrix}
\frac{n}{2}-1 \\ f_\ua
\end{bmatrix}
\begin{bmatrix}
\frac{n}{2}-1 \\ f_\da
\end{bmatrix}
\begin{bmatrix}
\frac{N-f_\ua-f_\da}{2}+n \\n
\end{bmatrix} \ .
\end{split}
\end{equation}



\begin{thebibliography}{99}


\bibitem{laughlin83}
R. B.  Laughlin,  Phys. Rev. Lett. {\bf 50}, 1395 (1983).

\bibitem{k09}
A.~Kitaev,
{\it Periodic table for topological insulators and superconductors},
AIP Conf. Proc. {\bf 1134} 22 (2009); arXiv:0901.2686.


\bibitem{srf09}
A.P.~Schnyder, S.~Ryu, A.~Furusaki, A.W.W.~Ludwig,
{\it Classification of Topological Insulators and Superconductors},
AIP Conf. Proc. {\bf 1134} 10 (2009); arXiv:0905.2029.


\bibitem{h83}
F.D.M.~Haldane,
Phys. Rev. Lett. {\bf 51}, 605 (1983).


\bibitem{asw84}
D.~Arovas, J.R.~Schrieffer, F.~Wilczek,
Phys. Rev. Lett. {\bf 53}, 722 (1984).


\bibitem{mr91}
G.~Moore, N.~Read,
Nucl. Phys. B {\bf360}, 362 (1991).

\bibitem{r09}
N.~Read,
Phys. Rev. B {\bf 79}, 045308 (2009).

\bibitem{bgn11}
P.~Bonderson, V.~Gurarie, C.~Nayak,
Phys. Rev. B {\bf 83}, 075303 (2011).

\bibitem{w95}
X.-G.~Wen,
Adv. Phys. {\bf 44}, 405 (1995).


\bibitem{es98}
R.A.J.~van~Elburg, K.~Schoutens,
Phys. Rev. {\bf B} 58, 15704 (1998).


\bibitem{abs01}
E.~Ardonne, P.~Bouwknegt, K.~Schoutens,
J. Stat. Phys. {\bf 102}, 421 (2001).


\bibitem{bbs10up}
B.~Estienne, B.A.~Bernevig, R.~Santachiara,
{\it Electron-Quasihole Duality and Second Order Differential Equation for Read-Rezayi
and Jacks Wavefunctions},
arXiv:1005.3475 (unpublished).


\bibitem{j89}
J.K.~Jain,
Phys. Rev. Lett. {\bf 63}, 199 (1989). 



\bibitem{bh08b}
B.A. Bernevig, F.D.M. Haldane,
Phys. Rev. Lett. {\bf 101}, 246806 (2008).


\bibitem{bh08c}
B.A. Bernevig, F.D.M. Haldane,
Phys. Rev. B {\bf 77}, 184502 (2008).


\bibitem{bh09}
B.A. Bernevig, F.D.M. Haldane,
Phys. Rev. Lett. {\bf 102}, 066802 (2009).

\bibitem{jj03}
G.S.~Jeon, J.K.~Jain,
Phys. Rev. B {\bf 68}, 165346 (2003).


\bibitem{hhr09}
T.H.~Hansson, M.~Hermanns, N.~Regnault, S.~Viefers,
Phys. Rev. Lett. {\bf 102}, 166805 (2009).


\bibitem{hhs09}
T.H.~Hansson, M.~Hermanns, S.~Viefers,
Phys. Rev. B {\bf 80}, 165330 (2009).


\bibitem{rxxup}
I.D.~Rodriguez, {\it et.al.}, 
{\it to be published}.


\bibitem{reg08}
N. Regnault, M.O. Goerbig, Th. Jolicoeur,
Phys. Rev. Lett. {\bf 101}, 066803 (2008).


\bibitem{mjv09}
M.V.~Milovanovic, Th.~Jolicoeur, I.~Vidanovic,
Phys. Rev. B {\bf 80}, 155324 (2009).


\bibitem{hrb11}
M.~Hermanns, N.~Regnault, B.A.~Bernevig, E.~Ardonne,
Phys. Rev. B {83}, 241302(R) (2011).


\bibitem{seidel2011}
A.~Seidel, K.~Yang,
{\it Gapless excitations in the Haldane-Rezayi state: The thin torus limit},
arXiv:1103.1903 (unpublished).



\bibitem{halperin83} 
B.I.~Halperin, 
Helv. Phys. Acta {\bf 56}, 75 (1983).


\bibitem{hr88}
F.D.M.~Haldane, E.H.~Rezayi,
Phys. Rev. Lett. {\bf 60}, 956 (1988).


\bibitem{readrezayi2}
N. Read, E.H. Rezayi, 
Phys. Rev. B {\bf 59}, 8084 (1999).


\bibitem{ymg88}
D.~Yoshioka, A.H.~MacDonald, S.M.~Girvin,
Phys. Rev. B {\bf 38}, 3636(R) (1988).


\bibitem{src07}
S.H.~Simon, E.H.~Rezayi, N.R.~Cooper,
Phys. Rev. B {\bf 75}, 075318 (2007).


\bibitem{thesis:green01}
D.~Green,
{\it Strongly Correlated States in Low Dimensions},
Ph.D. thesis, Yale University, New Haven (2001);
arXiv:cond-mat/0202455.


\bibitem{bh08a}
B.A. Bernevig, F.D.M. Haldane,
Phys. Rev. Lett. {\bf 100}, 246802 (2008).



\bibitem{fjm02}
B.~Feigin, M.~Jimbo, T.~Miwa, E.~Mukhin,
Int. Math. Res. Notices {\bf 23}, 1223 (2002).


\bibitem{bk05}
E.J.~Bergholtz, A.~Karlhede,
Phys. Rev. Lett. {\bf 94}, 026802 (2005).


\bibitem{bkw06}
E.J.~Bergholtz, J.~Kailasvuori, E.~Wikberg, T.H.~Hansson, A.~Karlhede,
Phys. Rev. B {\bf 74}, 081308(R) (2006).


\bibitem{sl06}
A.~Seidel, D.-H.~Lee,
Phys. Rev. Lett. {\bf 97}, 056804 (2006).


\bibitem{abk08}
E.~Ardonne, E.J.~Bergholtz, J.~Kailasvuori, E.~Wikberg,
J. Stat. Mech. P04016 (2008).


\bibitem{rr96}
N.~Read, E.~Rezayi,
Phys. Rev. B {\bf 54}, 16864 (1996).


\bibitem{gr00}
V.~Gurarie, E.~Rezayi,
Phys. Rev. {\bf B} 61, 5473 (2000).


\bibitem{arr01}
E.~Ardonne, N.~Read, E.~Rezayi, K.~Schoutens,
Nucl. Phys. B {\bf 607}, 549 (2001).


\bibitem{a02}
E.~Ardonne,
J. Phys. A {\bf 35}, 447 (2002).


\bibitem{sy08}
A.~Seidel, K.~Yang,
Phys. Rev. Lett. {\bf 101}, 036804 (2008).


\bibitem{as99}
E.~Ardonne, K.~Schoutens,
Phys. Rev. Lett. {\bf 82}, 5096 (1999).


\bibitem{all02}
E.~Ardonne, F.J.M.~van~Lankvelt, A.W.W.~Ludwig, K.~Schoutens,
Phys. Rev. B {\bf 65}, 041305 (2002).


\bibitem{h91}
F.D.M.~Haldane,
Phys. Rev. Lett. {\bf 67}, 937 (1991).

\bibitem{r32}
G. Rumer,
{\it Zur Theorie der Spinvalenz},
Nachr. d. Ges. d. Wiss. zu G\"{o}ttingen, M. P. Klasse, p. 337 (1932).

\bibitem{p33}
L. Pauling,
{\it The Calculation of Matrix Elements for Lewis Electronic Structures of Molecules},
J. Chem. Phys. {\bf 1}, 280 (1933).

\bibitem{cgt01}
A.~Cappelli, L.S.~Georgiev, I.T.~Todorov,
Nucl. Phys. B {\bf 599}, 499 (2001).

\bibitem{sal02}
K.~Schoutens, E.~Ardonne, F.J.M.~van~Lankvelt,
in Statistical Field Theories (NATO Science Series II: Mathematics, Physics and Chemistry),
A. Cappelli, G. Mussardo (editors),
p. 305-316, Kluwer Academic (2002);

\bibitem{rg00}
N.~Read, D.~Green,
Phys. Rev. B {\bf 61} 10267 (2000).

\bibitem{ww94}
X.-G.~Wen, Y.-S.~Wu,
Nucl. Phys. B {\bf 419}, 455 (1994).



\bibitem{mr96}
M.~Milovanovi\`{c}, N.~Read,
Phys. Rev. B {\bf 53}, 13559 (1996).



\bibitem{gfn97}
V.~Gurarie, M.~Flohr, C.~Nayak,
Nucl. Phys. B {\bf 498}, 513 (1997).



\bibitem{mjr08}
G.~Moller, Th.~Jolicoeur, N.~Regnault,
Phys. Rev. A {\bf 79}, 033609 (2009).



\bibitem{thesis:lankvelt04}
F.J.M.~van~Lankvelt,
{\it Quantum Hall spin Liquids},
PhD. thesis, University of Amsterdam (2004).


\bibitem{ds11}
S.~Davenport, S.H.~Simon, 
{\it to be published}. 


\bibitem{hms11up}
L.~Hormozi, G. M\"{o}ller, S.H.~Simon,
{\it Fractional quantum Hall effect of lattice bosons near commensurate flux},
arXiv:1109.3434 (unpublished).



\bibitem{li2008}
H.~Li, F.D.M.~Haldane,
Phys. Rev. Lett {\bf 101}, 010504 (2008).


\bibitem{sterdyniak2011}
A.~Sterdyniak, N.~Regnault, B.A.~Bernevig,
Phys. Rev. Lett {\bf 106}, 100405 (2011).


\bibitem{thomale2010}
R.~Thomale, B.~Estienne, N.~Regnault, B.A.~Bernevig,
{\it Decomposition of fractional quantum Hall states: New symmetries and approximations},
arxiv:1010.4837 (unpublished).


\bibitem{bw08a}
X.-G. Wen, Z. Wang
Phys. Rev. B {\bf 78}, 155109 (2008).


\bibitem{bw08b}
X.-G. Wen, Z. Wang,
Phys. Rev. B {\bf 77}, 235108 (2008).

\bibitem{bw10}
M. Barkeshli, X.-G. Wen,
Phys. Rev. B {\bf 82}, 245301 (2010).

\bibitem{bw10up}
M. Barkeshli, X.-G. Wen,
{\it Bilayer quantum Hall phase transitions and the orbifold non-Abelian fractional quantum
Hall states},
arXiv:1010.4270 (unpublished).

\bibitem{s10}
A.~Seidel,
Phys. Rev. Lett. {\bf 105}, 026802 (2010).

\bibitem{be11}
B. Estienne, B.A. Bernevig,
{\it Spin-Singlet Quantum Hall States and Jack Polynomials with a Prescribed Symmetry},
arXiv:1107.2534 (unpublished).


\bibitem{ers10}
B. Estienne, N. Regnault, R. Santachiara
Nucl. Phys. B {\bf 824}, 539 (2010). 

\bibitem{lm62}
E.~Lieb, D.~Mattis,
Phys. Rev. {\bf 125}, 164 (1962).

\bibitem{packages}
The packages are available for download from arXiv,
http://arxiv.org/src/1107.2232/anc.


\end{thebibliography}
\end{document}